\documentclass[12pt]{article}
\usepackage[utf8]{inputenc}
\usepackage{graphicx}
\usepackage{color}
\usepackage{amsfonts}
\usepackage{tabularx}
\usepackage{booktabs}
\usepackage{adjustbox} 
\usepackage{authblk}
\usepackage{longtable}
\usepackage{amsmath}
\usepackage{relsize}
\usepackage{tabularx} 
\usepackage[a4paper, margin=1in]{geometry}
\usepackage{setspace}
\usepackage{hyperref}
\usepackage{cleveref}
\usepackage{microtype}
\crefname{figure}{Web Figure}{Web Figures}
\Crefname{figure}{Web Figure}{Web Figures}
\usepackage[
    backend=biber,
    style=authoryear-comp,
    citetracker=true,
    uniquename=false,
    uniquelist=false,
    maxcitenames=2,
    mincitenames=1,
    maxbibnames=6,
    minbibnames=6,
    sorting=nyt
]{biblatex}

\AtEveryCitekey{%
  \ifciteseen
    {\setcounter{maxnames}{1}}%
    {%
      \ifnum\value{author}<4
        \setcounter{maxnames}{\value{author}}%
      \else
        \setcounter{maxnames}{3}%
      \fi
    }%
}

\title{Meta-Analysis of High-Dimensional Surrogate Markers}

\author[1,2]{Arthur Hughes}
\author[1,2,3]{Rodolphe Thiébaut}
\author[4]{Layla Parast}
\author[1,2,*]{Boris P. Hejblum}

\affil[1]{INSERM Bordeaux Population Health Research Center, INRIA SISTM, University of Bordeaux, F-33000 Bordeaux, France}
\affil[2]{Vaccine Research Institute, F-94000 Créteil, France}
\affil[3]{Centre Hospitalier Universitaire de Bordeaux, Service d’Information Médicale, INSERM, Bordeaux F-33000, France}
\affil[4]{Department of Statistics and Data Science, University of Texas at Austin, Austin, TX 78712, United States of America}
\affil[*]{\textbf{Corresponding author:} Boris P. Hejblum, 
Bordeaux Population Health, Université de Bordeaux, 146 Rue Léo Saignat 33076 Bordeaux, France. Email: \texttt{boris.hejblum@u-bordeaux.fr}}

\date{} 
\begin{document}

\begin{refsection}

\maketitle

\begin{abstract}
When direct measurement of a clinically relevant primary endpoint in a clinical trial is infeasible, a surrogate endpoint may be used instead to infer treatment effects. Trial-level surrogates predict the average treatment effect on the primary endpoint and may be evaluated within the meta-analytic framework. However, traditional methods are ill-suited to the complex high-dimensional data now increasingly collected in modern trials, such as omics data. Although methods for high-dimensional surrogate evaluation exist, they have largely been developed for single-trial settings and therefore cannot assess surrogate generalisability. Here, we propose RISE-Meta, an approach for evaluating trial-level surrogate markers in the multi-trial, high-dimensional setting. In the first stage, an existing nonparametric method is applied to individual participant data to derive study-level surrogacy metrics for each candidate marker. Next, random-effects meta-analysis combines these metrics across studies, and equivalence testing provides operational criteria for surrogate validity. Finally, a subset of candidates is combined into a composite signature through a weighting scheme to improve surrogacy relative to any individual candidate. We evaluate RISE-Meta in both simulation studies and real data applications. In an application to high-dimensional data, we analyse gene expression as trial-level surrogate markers for the antibody response to seasonal influenza vaccination, while in a low-dimensional application we compare RISE-Meta to a reference meta-analytic approach and observe strong agreement between the two.
\end{abstract}

\vspace{1em} 

\newpage

\section{Introduction}\label{sec1}

The main objective of a clinical trial is to evaluate the effect of a treatment on a clinically relevant primary endpoint. However, direct measurement of the primary endpoint may not be feasible due to cost, time, or ethics. In such cases, one may instead rely on a surrogate marker, a variable used to infer treatment effects on the primary endpoint. For example, in trials evaluating the efficacy of yellow fever vaccination in European populations, relying on infection rates is problematic since yellow fever virus is not endemic to Europe. Treatment efficacy may be assessed instead using neutralising antibody titres, which serve as a surrogate marker for infection rates \parencite{Jonker2013}.

However, establishing the validity of a candidate surrogate has been a longstanding problem in biostatistics, and has been approached using multiple paradigms. Among these is the meta-analytic framework, which evaluates surrogate candidates using data from multiple clinical trials \parencite{Daniels1997, Buyse2000, Molenberghs2002, Alonso2004}. Where data from multiple trials are available, the meta-analytic framework is attractive as it exploits between-trial heterogeneity to assess whether surrogates predict treatment effects across clinical and demographic settings.

The historical development of the meta-analytic framework has clarified the conceptualisation of surrogate markers. \textcite{Buyse2000} introduced the distinction between \textit{individual-level surrogates} (ILS), associated with the primary endpoint at the participant level, and \textit{trial-level surrogates} (TLS), for which the average treatment effect on the surrogate predicts the average treatment effect on the primary endpoint across trials. Two metrics, $R^{2}_{\text{indiv}}$ and $R^{2}_{\text{trial}}$, were proposed to assess surrogacy at the individual- and trial-level, respectively. These distinctions are important as ILS and TLS differ both conceptually \parencite{Korn2016} and in their use cases \parencite{Alonso2005}. In this work, we focus on meta-analytic methods for evaluating trial-level surrogate markers.

Advances in the cost and accuracy of high-throughput technologies have enabled more routine generation of \textit{high-dimensional} biomarkers in modern biomedical studies. In such settings, the number of candidate markers can exceed the number of samples by several orders of magnitude. A typical example is gene expression profiling, where tens of thousands of genes are measured simultaneously in treated and control individuals. These data can characterise the molecular mechanisms underlying treatment effects and identify biomarkers predictive of the host response \parencite{Prachi2012}. Thus, high-dimensional markers are both practical and biologically relevant candidates for surrogate markers, but their scale and complexity pose challenges for evaluation.

Public data-sharing initiatives have generated an abundance of high-dimensional datasets for the same treatment, accessible through repositories such as the \textit{Cancer Genome Atlas}, \textit{Gene Expression Omnibus}, and \textit{ImmPort} \parencite{Edgar2002, Weinstein2013, Bhattacharya2018}. Although individual studies are often limited in sample size, their aggregation can produce large datasets. This naturally motivates the use of meta-analytic approaches, which provide a principled framework for combining information across studies. Such approaches have already been applied in high-dimensional settings outside of the context of surrogate evaluation \parencite{Choi2003, Tseng2012, Liyanage2024}.

However, although traditional meta-analytic approaches for surrogate evaluation are well established, they are not well suited to high-dimensional settings. A key limitation is their reliance on strong distributional assumptions for individual participant data (IPD). For example, the methods described by \textcite{Buyse2000} and \textcite{Molenberghs2002} assume a bivariate normal distribution for the surrogate and the primary endpoint. These assumptions become increasingly hard to verify and restrictive as the dimensionality and complexity of the data grow, limiting the applicability of such methods in many modern contexts.

The evaluation of high-dimensional surrogate markers is an emerging area of research, with only a few methods currently addressing this setting. \textcite{Agniel2022} developed methodology to estimate the proportion of treatment effect explained (PTE) by a group of markers but did not provide quantitative rules for selecting individual markers. \textcite{Zhou2022} used high-dimensional linear mediation models to estimate the PTE, but their method is not robust to model misspecification. \textcite{Hughes2025} introduced RISE, a two-stage, nonparametric approach which is robust to the distribution of the IPD and suitable for small sample sizes. A shared limitation of all these approaches is that they operate within the single-trial framework, precluding the evaluation of generalisability and effective use of the wealth of available public data.

In this work, we extend the work of \textcite{Hughes2025} by adapting RISE to the meta-analytic setting, enabling the evaluation of high-dimensional surrogate markers across multiple trials. The paper is structured as follows. Section \ref{sec2} summarises the existing nonparametric approaches for surrogacy evaluation and introduces RISE-Meta, our meta-analytic extension. Section \ref{sec3} presents a simulation study assessing its statistical properties. In Section \ref{sec4}, we apply RISE-Meta to real datasets. First, in a typical high-dimensional setting, we evaluate gene expression as a trial-level surrogate for the antibody response to seasonal influenza vaccination. Then, we apply RISE-Meta to classical low-dimensional datasets to compare the results with a reference meta-analytic approach introduced in \textcite{Buyse2000}. Section \ref{sec5} concludes with a discussion of strengths, limitations, and future research directions.

\section{Methods}\label{sec2}

\subsection{Notation and data structure}

\paragraph*{\textbf{Active versus control design}\\}

In this setting, we observe data from $m = 1,\dots,M$ trials, each with $i = 1,\dots,n_{m}$ individuals receiving the same binary treatment $A_{i,m} \in \{0,1\}$, where $1$ denotes active treatment and $0$ denotes control. Let $Y_{i,m}^{a}$ and $\boldsymbol{S}_{i,m}^{a} = (S_{i,1,m}^{a},\dots,S_{i,J,m}^{a})^{T}$ denote the primary endpoint and vector of $J$ common surrogate candidates, respectively, for individual $i$ in trial $m$ receiving treatment $a$.
Without loss of generality, we assume that larger values of $Y$ and $S_j$ correspond to better outcomes. In addition, we assume the trials are independent and that the data within each trial are independent and identically distributed. 

\paragraph*{\textbf{Paired design}\\}

Trials with high-dimensional data do not always follow a randomised treatment/control regime, but can instead consist of the longitudinal measurements of biomarkers in the same individuals before and after receipt of treatment. In this case, for each individual and a given post-treatment timepoint, we have paired measurements $\{Y_{i,m}^{0},Y_{i,m}^{1},\boldsymbol{S}_{i,m}^{0},\boldsymbol{S}_{i,m}^{1}\}$, where the superscripts $a = 0$ and $a = 1$ refer to pre- and post-treatment measurements, respectively.

\subsection{Existing nonparametric approaches for surrogate evaluation}\label{sec:existing}

In this subsection, we summarise existing nonparametric methods for evaluating surrogate markers. This framework was first proposed by \textcite{Parast2024} and later extended in \textcite{Hughes2025} to the high-dimensional setting.  
Since the discussion here focuses on the single-marker, single-trial context, we drop the subscripts $j$ and $m$, and denote an arbitrary surrogate candidate by $S$.

\subsubsection{Within-study surrogacy parameter}

 \textcite{Parast2024} first define average treatment effects on the primary endpoint and candidate $S$ in terms of U-statistics as follows: $$U_{Y} = \mathbb{P}\bigl(Y_{i}^{1} > Y_{l}^{0}\bigr) + \tfrac{1}{2}\mathbb{P}\bigl(Y_{i}^{1} = Y_{l}^{0}\bigr), \qquad U_{S} = \mathbb{P}\bigl(S_{i}^{1} > S_{l}^{0}\bigr) + \tfrac{1}{2}\mathbb{P}\bigl(S_{i}^{1} = S_{l}^{0}\bigr),$$
where $i$ and $l$ index independent observations from the treated and control arms. These measures, defined on the interval $\lbrack 0,1\rbrack$, represent the average probabilities that a randomly sampled treated individual has a greater value than a randomly sampled control for the variables $Y$ and $S$, respectively. We then define within-trial surrogacy as the difference between these effects: 
$$\delta = U_{Y} - U_{S},$$ 
which may be interpreted as the bias in estimating the average treatment effect on $Y$ through observation of $S$ alone. The closer this bias is to 0, the better $S$ is as a trial-level surrogate in this single-trial framework. These quantities may be estimated nonparametrically using rank-based methods, with associated confidence intervals (CIs) estimated using U-statistic theory.

\textcite{Hughes2025} proposed that a valid surrogate in this framework may be determined by testing the hypotheses 
$$H_{0}: \delta \notin \bigl(-\varepsilon,\varepsilon \bigr), \qquad H_{1}: \delta \in \bigl(-\varepsilon,\varepsilon \bigr),$$
where $\varepsilon$ is some small, pre-determined constant, called the \textit{equivalence bound}, defining the \textit{equivalence margin} $\bigl(-\varepsilon,\varepsilon \bigr)$. This bound can either be chosen \textit{a priori} or through the data-driven approach described in \textcite{Parast2024}, whereby the investigator specifies a target power to detect a treatment effect on $Y$ via $S$.

Equivalence is checked with a two one-sided test (TOST) procedure \parencite{Schuirmann1987}, which must reject both that $\delta \leq -\varepsilon$ and $\delta \geq \varepsilon$. Further detail on estimation, as well as a set of assumptions needed to avoid a \textit{surrogate paradox} situation, are provided in the original publications \parencite{Hughes2025, Parast2024} and in the Supplementary Material of this article.

\subsection{Overview of RISE-Meta}\label{sec:overview}

First, we provide a broad overview of our proposed method, RISE-Meta, which is an extension of the nonparametric approach to the meta-analytic setting. Let $\widehat{\delta}_{j,m}$ refer to the estimate of $\delta$ for marker $j$ in study $m$. RISE-Meta consists of three sequential steps:
\begin{enumerate}
    \item \textbf{Within-trial screening across candidates:} for each trial, estimate marker-specific surrogacy metrics $\widehat{\delta}_{j,m}$ and their sampling error, as described in Section~\ref{sec:existing}.
    \item \textbf{Meta-analysis of within-trial effects for each candidate:} pool $\widehat{\delta}_{j,m}$ across studies using random-effects meta-analysis. Perform equivalence testing on the pooled effects and adjust all p-values to control for the overall false discovery rate. Use adjusted p-values to define a set of promising markers. 
    \item \textbf{Construction and evaluation of a surrogate signature}: combine the set of identified promising markers into a single composite signature using a weighted sum, where markers with desirable surrogacy properties are given larger weights. Evaluate the trial-level surrogacy of the signature using hypothesis testing, prediction intervals, and concordance-based metrics. 
\end{enumerate}
A graphical depiction of these three steps is given in the Supplementary Material. In order to avoid problems associated with post-selection inference and overfitting, stages 2 and 3 should be applied to different data \parencite{Ball2020}. This could be done either through within-study sample splitting or, when enough studies are available, retaining a small number of studies for the third stage only. 

\subsection{Random-effects meta-analysis model}\label{sec:model}

In this section, we present the random-effects meta-analysis model used to pool effects $\widehat{\delta}_{j,m}$ across studies, as well as our recommended estimation procedure. We note that there is considerable debate about the merits of the random and fixed-effect approaches to meta-analysis in the literature \parencite{McKenzie2024}. While throughout this section we describe the random-effects approach, this is easily reduced to a fixed-effect approach, and we include this as an option in the \texttt{R} package implementing these methods, \texttt{SurrogateRank}.

\subsubsection{Model specification}

We assume a hierarchical model for the estimates of the within-study effects for the $j$th marker. In the first level, we assume that the true effects follow $$\delta_{m,j} \sim N(\mu_{j}, \tau^{2}_{j}),$$
with fixed mean $\mu_{j}$ and variance $\tau^{2}_{j}$, which we refer to as the \textit{between-study variance}. In the second level, we assume that the estimates of these values $\widehat{\delta}_{m,j}$ are normally distributed around the true parameter values  $$\widehat{\delta}_{m,j} \sim N(\delta_{m,j}, \sigma^{2}_{m,j}),$$ where the variance $\sigma^{2}_{m,j}$ is referred to as the \textit{within-study variance}. It is worth emphasising here that these normality assumptions are imposed only on study-level quantities. In contrast to classical meta-analytic approaches, we do not assume any parametric form (such as bivariate normality of $Y$ and $S$) for the IPD within studies.

\subsubsection{Estimation}

The within-study variances $\sigma^{2}_{m,j}$ are estimated within each study, and the between-study variance $\tau^{2}_{j}$ is estimated through a numerical algorithm implementing restricted maximum likelihood.
The pooled effect $\mu_{j}$ is classically estimated using a weighted sum $$\widehat{\mu}_{j}=\dfrac{\sum\limits_{m = 1}^{M}\widehat{w}_{m,j}\widehat{\delta}_{m,j}}{\sum\limits_{m = 1}^{M}\widehat{w}_{m,j}}$$

\noindent where the weights are the inverse of the total variance within each study, i.e. $\widehat{w}_{m,j} = (\widehat{\sigma}^{2}_{m,j} + \widehat{\tau}^{2}_{j})^{-1}$.
The standard error, $s_{j}$, of this pooled estimate may be estimated through the Hartung-Knapp-Sidik-Jonkman (HKSJ) approach, which is more robust than the conventional estimator, especially when the number of studies $M$ is small \parencite{Hartung2001, IntHout2014}. Then, $(1-\alpha)\times 100\%$ CIs can be constructed as $\widehat{\mu}_{j} \pm q_{M-1, 1 - \alpha/2} \times \widehat{s}_{j}$, where $q_{M-1, 1 - \alpha/2}$ is the $(1 - \alpha/2)\times 100\%$ quantile of the $t$-distribution with $M-1$ degrees of freedom. Further details for estimation are given in the Supplementary Material. 

\subsection{Hypothesis testing}

Next, we perform hypothesis tests on the pooled effect sizes $\widehat{\mu}_{j}$ to make operational conclusions about surrogate validity. Again, we adopt the framework for equivalence testing using the two one-sided test (TOST) procedure \parencite{Schuirmann1987} briefly described in Section \ref{sec:existing}. Specifically, this requires testing two null hypothesis, which we term upper (U) and lower (L): $$H_{0,j}^{L}: \mu_{j} \le -\varepsilon, \quad
H_{1,j}^{L}: \mu_{j} > -\varepsilon,$$ $$H_{0,j}^{U}: \mu_{j} \ge \varepsilon, \quad
H_{1,j}^{U}: \mu_{j} < \varepsilon,$$ and concluding $\mu_{j} \in (-\varepsilon, \varepsilon)$ only if both tests are rejected. Using the HKSJ method, p-values for the lower and upper hypotheses can be estimated as $\widehat{p}_{j}^{L} =1 - F_{M-1}\left (\dfrac{\widehat{\mu}_{j} + \varepsilon}{\widehat{s}_{j}} \right)$ and $\widehat{p}_{j}^{U} = F_{M-1}\left (\dfrac{\widehat{\mu}_{j} - \varepsilon}{\widehat{s}_{j}} \right)$, respectively, where $F_{M-1}(.)$ denotes the cumulative distribution function of the $t$-distribution with $M-1$ degrees of freedom. The p-value for the TOST is defined as $\widehat{p}_{j} = \max(\widehat{p}_{j}^{L}, \widehat{p}_{j}^{U})$, where a significant p-value at the $\alpha$-level implies that the $(1-2\alpha)\times 100\%$ CI of $\widehat{\mu}_{j}$ lies within $(-\varepsilon, \varepsilon)$. 

Finally, due to the large number of hypothesis tests performed, we control the false discovery rate using the Benjamini-Hochberg procedure \parencite{Benjamini1995}, and refer to the resulting p-values as the adjusted p-values, denoted $\widehat{p}_{j}^{\,*}$.

\paragraph{Additional evaluation metrics.}

Although CIs and p-values provide useful metrics for operational decision making in the context of high-dimensional surrogate candidates, it may be unwise to uniquely rely on these criteria. In particular, prior work has shown that CIs may be misleading in the presence of substantial between-study variability \parencite{IntHout2016}. This has led to the advocacy of prediction intervals (PIs), which represent the likely range of effect estimates in new, similar trials. We argue that PIs are highly relevant in the context of surrogate markers, since the goal is to use the surrogate to predict the treatment effect on the primary endpoint in a new trial. Using the HKSJ approach, a $(1-\alpha)\times 100\%$ PI can be estimated as $\widehat{\mu}_{j} \pm q_{M-1, 1 - \alpha/2} \times \sqrt{\widehat{s}^{2}_{j} + \widehat{\tau}_{j}^{2}}$.

In addition, we use the concordance correlation coefficient (CCC) \parencite{Lin1989} between the treatment effects on $Y$, $\boldsymbol{\widehat{U}_{Y}} = (\widehat{U}_{Y,1},\dots,\widehat{U}_{Y,M})^{T}$, and the treatment effects on $\boldsymbol{S}$, $\boldsymbol{\widehat{U}_{S}} = (\widehat{U}_{S,1},\dots,\widehat{U}_{S,M})^{T}$, as a further validation metric for trial-level surrogacy. Like other correlation coefficients, this sits within the interval $\lbrack -1, 1\rbrack$, with similar interpretation. However, this metric also penalises differences from the identity line, allowing for a more nuanced evaluation in the context of our treatment effects, which are on a standardised scale. Further details are given in the Supplementary Material. 

\subsection{Construction of a composite signature}

The final stage of the method involves combining the promising surrogate candidates together to form a composite signature. First, we define a set of indices of the significant markers from the previous tests $\Gamma = \{j: \widehat{p}_{j}^{\,*} < \alpha \}$. We combine the markers defined by this set together using a weighted sum $$\boldsymbol{\gamma}_{\Gamma} = \sum_{j \in \Gamma} \widehat{\lambda}_j\, \boldsymbol{\bar{S}}_j,$$where $\boldsymbol{\bar{S}}_j$ is the vector of the $j$th surrogate standardised to have 0 mean and 1 variance, and $\widehat{\lambda}_j$ is the estimated \textit{signature weight} for the $j$th marker, which we describe below. 

The signature weights should be chosen in order to increase the prominence of markers with better surrogacy properties. We consider two such properties, which combine together to make the final signature weight: the strength of the surrogate, and its consistency across trials.
The first component depends on the pooled mean of the $j$th candidate, and is formed as $\widehat{a}_j = \frac{\varepsilon - |\widehat{\mu}_{j}|}{\varepsilon}$, which gives maximum weight when $\widehat{\mu}_{j} = 0$ and minimum weight when $\widehat{\mu}_{j}$ is on the boundary $\varepsilon$. This component adapts the idea behind the weighting scheme presented in \textcite{Hughes2025} to be more stable for small values of $\widehat{\mu}_{j}$. The second component concerns the precision of estimation within and between-trials, determined by the sum of the total variances for each trial $\widehat{b}_j = \sum_{m=1}^M(\widehat{\sigma}_{m,j}^2 + \widehat{\tau}_{j}^2)^{-1}$.  

We then construct the final signature weights as a standardised combination of the two components:$$\widehat{\lambda}_j = \frac{\widehat{a}_j\widehat{b}_j}{\max_{j \in \Gamma} \{\widehat{a}_j\widehat{b}_j\}}.$$

\section{Simulation study}\label{sec3}

\subsection{Simulation setup}

We first use a parametric data generating process to evaluate the meta-analysis equivalence testing procedure proposed in Section \ref{sec2}.
Our primary objectives are to assess the calibration of the test, via the empirical false positive rate (FPR), and its empirical power. The calibration assessment uses only invalid surrogates, while the power assessment uses only valid surrogates. 

Valid surrogates in our framework are defined as those for which $\mu_j \in (-\varepsilon, \varepsilon)$, and invalid surrogates otherwise. As the null hypothesis is composite, its definition corresponds to a region of the parameter space. To assess calibration, it is therefore natural to consider the \textit{least favourable configuration} (LFC), in which invalid surrogates lie on the boundary of the null, i.e. $\mu_j \in \{-\varepsilon, \varepsilon\}$. If the empirical FPR is controlled at level $\alpha$ in this setting, it will also be controlled for more favourable configurations.

We consider two main scenarios corresponding to the evaluation of calibration and power. Throughout, we generate results across $J = 100,000$ independent marker draws, with nominal FPR $\alpha = 0.05$ and equivalence bound $\varepsilon = 0.1$.
In Scenario 1, we generate only invalid surrogates under the LFC to assess the empirical FPR. Within this scenario, we examine results at various values of $M$, the number of studies, ($M=3,10,25$), and $n_m$, the sample size within each study ($n_m=10,50,250$). In addition, we examine results at varying levels of between-study and within-study variability, with fixed $M=n_m=20$. To balance effects on either side of the null boundary, we set $\mu_j = -\varepsilon$ and $\mu_j = \varepsilon$ with equal probability. In Scenario 2, we generate only valid surrogates to assess power, with $\mu_j$ allowed to vary across the valid region as $\mu_j \sim \mathcal{U}(-\varepsilon, \varepsilon)$. Within this scenario, we examine results at various values of $M$ ($M=3,10,25$), and varying levels of between-study and within-study variability, with fixed $n_m=50$.

For each marker $j = 1,\dots,J$, we first draw the true mean $\mu_j$ based on the scenarios below, which depend on if we want a valid or invalid surrogate. We then simulate marker-specific between-study heterogeneity as $\tau_j^2 \sim \mathcal{U}(0, u_{\tau^{2}, max})$. Conditional on these parameters, the true study-specific effects are generated as $$
\delta_{m,j} \sim \mathcal{N}(\mu_j, \tau_j^2), \qquad m = 1,\dots,M.
$$

At the second stage, we define the within-study variance as $\sigma_{m,j}^2 = \nu_j / n_{m}$, where $\nu_j \sim \mathcal{U}(0, u_{\nu,\max})$ introduces marker-specific variability. The observed effect estimates are then generated as $$
\widehat{\delta}_{m,j} \sim \mathcal{N}(\delta_{m,j}, \sigma_{m,j}^2).
$$

\subsection{Results}

\subsubsection{Scenario 1: calibration}

In scenario 1, where only invalid surrogates are generated, we investigated the calibration of the statistical test. The results are presented as calibration plots showing empirical against nominal FPR. 

\begin{figure}[htbp]
    \centering
    \includegraphics[width=\textwidth]{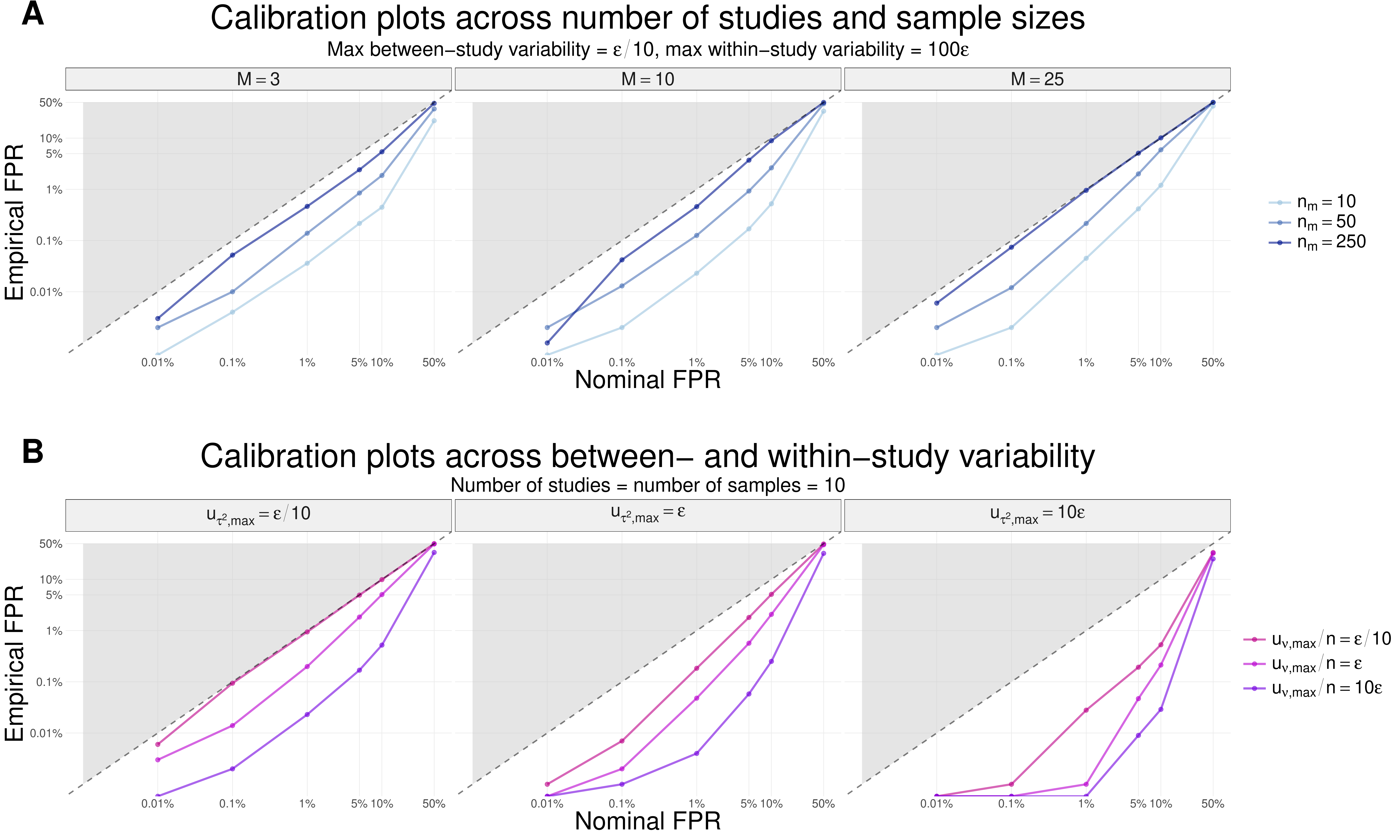}
    \caption{\textbf{Calibration plots comparing nominal and empirical false positive rates across A) samples and B) variability settings}. Each panel shows the nominal FPR $\alpha$ versus the empirical FPR, computed as the mean over 100,000 independent simulations for each configuration. The dashed diagonal represents perfect calibration. The shaded upper triangle delimits anti-conservative behaviour and the lower triangle conservative behaviour. The equivalence bound was set to $\varepsilon = 0.1$. In \textbf{A}), the number of studies $M$ (facets) and within-study sample size $n_{m}$ (lines) were varied, and the heterogeneity parameters were fixed at $u_{\tau^{2}, max} = \varepsilon/10$ and $u_{\nu,\max} = 100\varepsilon$, whereas in \textbf{B}) the between-study variability $u_{\tau^{2}, max}$ (facets) and within-study variability $u_{\nu,\max}/n$ (lines) were varied, and the sample parameters were fixed at $M = n_{m} = 10$. \textbf{Alt text:} Plots showing the claimed false positive rate against the observed false positive rate in simulations across various settings.}
    \label{fig:fig1}
\end{figure}

Figure \ref{fig:fig1}A shows calibration plots for number of studies $M = 3,\, 10,\, 25$, within-study sample size $n_{m} = 10,\, 50,\, 250$, and fixed maximum heterogeneity parameters $u_{\tau^{2}, max} = \varepsilon/10$, $u_{\nu,\max} = 100\varepsilon$. Figure \ref{fig:fig1}B shows calibration plots for maximum between-study variability $u_{\tau^{2}, max} = \varepsilon/10,\, \varepsilon,\, 10\varepsilon$ and max within-study variability $u_{\nu, max}/n_{m} = \varepsilon/10,\, \varepsilon,\, 10\varepsilon$ at fixed sample parameters $M = n_{m} = 10$.

In all cases, the empirical FPR was bounded by the claimed FPR, providing confidence about the test's statistical validity even when estimation is based off a small number of samples or studies, or in presence of extreme heterogeneity. As expected, calibration improves with increasing sample sizes, number of studies, and with decreasing heterogeneity, with near perfect calibration being achieved for the most favourable combinations of these values. Overall, these results indicate a conservative and valid test procedure, which is desirable in the context of surrogate marker evaluation, where an inflated false positive rate could have serious consequences. 

\subsubsection{Scenario 2: power}

\begin{figure}[htbp]
    \centering
    \includegraphics[width=\textwidth]{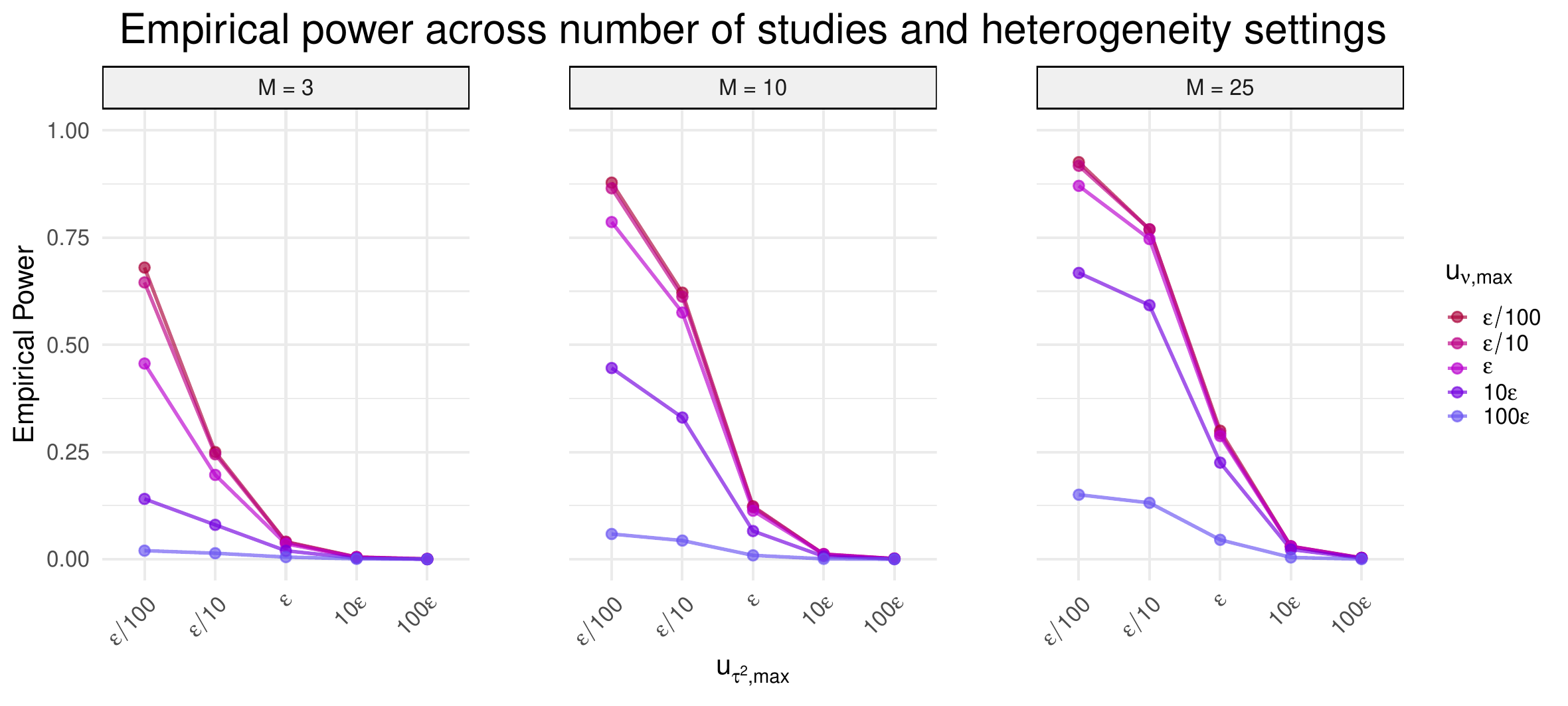}
    \caption{\textbf{Empirical power across different heterogeneity settings and number of studies}. The parameters $u_{\tau^{2}, max}$ and $u_{\nu,\max}$ control the maximum between-study and within-study heterogeneity, respectively. Both are expressed relative to the equivalence bound $\varepsilon = 0.1$. The within-study sample size was fixed at 50. \textbf{Alt text:} Graphs showing the observed power against the between-study heterogeneity.}
    \label{fig:fig2}
\end{figure}

In scenario 2, where only valid surrogates are generated, we investigated the power of the statistical test across settings. Figure \ref{fig:fig2} shows the empirical power as a function of maximum between- and within-study variability $u_{\tau^{2},max} = u_{\nu, max} = \{\varepsilon/100,\, \varepsilon/10,\, \varepsilon,\, 10\varepsilon,\, 100\varepsilon \}$ for number of studies $M = 3,\,10,\,25$ at a fixed sample size of $n_{m} = 50$. As expected, the power is greatest when the number of studies is high and the heterogeneity is low. The heterogeneity parameters have a greater impact on the empirical power than the number of studies, and in the cases of extremely high heterogeneity, the power to detect true positives reduces to 0.

\subsection{Supplementary simulations}

Although the parametric simulation allows fine control of parameters, it is also important to examine validity when the model may not hold. This can be done with nonparametric simulation, generating null datasets by random permutation of a real dataset while preserving its realistic properties. The results of this nonparametric simulation study, as well as further results from the parametric simulation study, are presented in the Supplementary Material. These results further support the validity of the test procedure and provide justification for our recommended analysis choices. 

\section{Data applications}\label{sec4}

\subsection{High-dimensional data example}

We applied RISE-Meta to public multi-trial data on trivalent inactivated influenza vaccine (TIV) to evaluate high-dimensional trial-level surrogates of the downstream immune response. These data were accessed through the ImmuneSpace platform (\url{immunespace.org}) as part of the Immune Signatures 2 (IS2) project \parencite{DirayArce2022}. 
TIV is an inactivated vaccine designed to induce antibodies against three circulating influenza strains. The primary endpoint is the antibody response to TIV, defined as the mean of the strain-specific antibody responses across the three strains targeted by each vaccine formulation four weeks after vaccination. Strain-specific antibody titres were measured using either neutralising antibody (nAb) or haemagglutination inhibition (HAI) assay, with nAb prioritised. 

Initially, the surrogate candidates consisted of the expression of 10,086 genes sequenced from either whole blood or peripheral blood mononuclear cell samples measured the day after vaccination. Rather than working at the gene-level, we aggregated these measurements to the geneset-level, which are predefined groups of functionally related genes. This transformation provides an initial dimension reduction, thereby reducing the multiple testing burden, and improves interpretability of the results. We used the Blood Transcriptional Modules (BTMs) \parencite{Li2013} as predefined genesets and summarised each module by the mean expression of its constituent genes. After filtering for modules with descriptive titles, we obtained $J = 258$ input features. Further details, as well as an evaluation of the impact of these choices, are provided in the Supplementary Material.

Each study involved paired data, with the primary endpoint and candidate surrogates measured before and after vaccination for each subject. Antibody responses were measured at days 0 and 28, where we allowed $\pm 7$ days for the post-vaccination measurement to account for minor differences in study design. Gene expression was measured at days 0 and 1. After filtering studies with fewer than 5 participants and individuals missing pre- or post-vaccination measurements, the data contained paired measurements from 334 individuals across 4 studies. We used 66\% of the data within each study for the marker-level meta-analysis, and reserved the remaining data for evaluation of the composite signature. The unique study identifiers in ImmPort (\url{immport.org}) \parencite{Bhattacharya2018} and IS2, and their corresponding numbers of individuals, are SDY56 ($n = 58$), SDY80 ($n = 56$), SDY180 ($n = 12$), and SDY1276 ($n = 208$). A full description of the IS2 TIV datasets and the data preprocessing steps for this analysis are given in the Supplementary Material.
The significance level was set to $\alpha = 0.05$, and the Benjamini-Hochberg (BH) procedure was applied to adjust the p-values for multiple testing. The value of $\varepsilon$ was chosen using the power approach described in \textcite{Parast2024}, taking the mean of $\varepsilon$ across studies such that within each study there is $80\%$ power to detect a treatment effect on the antibodies based on observation of the surrogate candidates. This gave $\varepsilon \approx 0.205$ in the screening stage and $\varepsilon \approx 0.148$ for the evaluation of the composite signature.

The results of the marker-level meta-analysis screening stage are shown in Figure \ref{fig:fig4}: seven markers remained significant after multiplicity correction. Their point estimates are positive and close to 0, suggesting a small underestimation of the treatment effect. Many of the top pathways are related to interferon/antiviral signalling, dendritic cell/antigen presentation, and monocyte signatures, giving these results a coherent immunological interpretation. We return to this point in the Discussion.
The results of the evaluation of the composite 7-marker signature on independent data are shown in Figure \ref{fig:fig5}. The estimates of $\delta_{m}$ are close to 0 across trials, with minor differences in point estimates and standard errors. The pooled effect is close to 0, and its $90\%$ CI lies within the equivalence margin, with $p = 0.004$ from the TOST. The PI is also small, reflecting minimal between-trial variability, with $\tau^{2} \approx 6 \times 10^{-11}$. In addition, the CCC is 0.94, indicating strong concordance between treatment effects on the primary endpoint and on this 7-marker signature. 
Together, these results illustrate that RISE-Meta is capable of deriving strong and interpretable trial-level surrogates in high-dimensional experiments. 

\subsection{Low-dimensional data example}

In addition, we applied RISE-Meta to two low-dimensional datasets to compare it with a reference method for meta-analytic surrogate evaluation. This is the only setting in which a direct comparison with classical approaches is possible, as these methods cannot be extended to high-dimensional data without substantial modification, which RISE-Meta is designed to address. Agreement in this setting would support the use of RISE-Meta in classical data settings where investigators desire to avoid strong distributional assumptions on the IPD.

The reference approach, introduced in \textcite{Buyse2000} and described in full across multiple sources \parencite{Molenberghs2002, Tibaldi2003, 2005}, is referred to as the \textit{bivariate joint modelling} approach. These reference approaches and the public datasets we use as examples in this section are accessible from the \texttt{Surrogate} package in \texttt{R} \parencite{VanDerElst2014}. Throughout, due to numerical convergence issues, we were unable to fit full or reduced bivariate mixed-effect models to the data, and opted instead to fit full bivariate fixed-effect models through the \texttt{BifixedContCont()} function. In the Supplementary Material, we give a more detailed description of this procedure. Since the low-dimensional datasets consisted of a single surrogate candidate each, the application of RISE-Meta only involves steps 1 and 2 described in Section~\ref{sec:overview}. For TOST equivalence testing, we used equivalence bound $\varepsilon = 0.1$ and $\alpha = 0.05$.

The first dataset, \texttt{ARMD}, comes from a clinical trial of patients with age-related macular degeneration treated with interferon-$\alpha$ or placebo. The primary endpoint is the change in visual acuity (CVA) at 12 months, and the candidate surrogate is CVA at 6 months. Although these data come from a single trial, we follow previous applications in the literature and use the treatment center as the unit of analysis \parencite{Molenberghs2002}. Further discussion on this choice is provided in the Supplementary Material and in \textcite{Abrahantes2004}. After filtering to centers with at least 5 total patients and at least 2 patients in each treatment group, we retained 109 patients across 15 centers. Filtering the smallest units was necessary to ensure non-degenerate estimation of the within-trial variance $\sigma_{m}^{2}$ when using RISE-Meta. 

The second dataset, \texttt{Ovarian}, comes from 4 trials in patients with advanced ovarian cancer comparing cyclophosphamide plus adriamycin plus cisplatin with cyclophosphamide plus cisplatin. The primary endpoint is the log of the overall survival time (OS), and the candidate surrogate is the log of the progression-free survival time (PFS). Again, we take the treatment center as the unit of analysis. After applying the same filtering criteria, we retained 1,168 patients across 42 centers.

To compare the methods fairly, we report both the $R^{2}_{trial}$, the CCC, and their 95$\%$ CIs (estimation detail in Supplementary Material). Figure~\ref{fig:fig3} shows a graphical summary of these results, indicating strong agreement between the two methods for both datasets. In particular, both methods conclude a moderately strong surrogate in the \texttt{ARMD} data and a strong surrogate in the \texttt{Ovarian} data in the context of their respective treatment comparisons. In addition, RISE-Meta gave significant p-values ($0.0067$ for \texttt{ARMD} and $7\times10^{-24}$ for \texttt{Ovarian}), indicating evidence for trial-level surrogacy. Overall, this example suggests RISE-Meta agrees with the reference method in low-dimensional settings where the normality assumption is more likely to hold.

\section{Discussion}\label{sec5}

In this article, we introduced RISE-Meta, a multi-stage framework for evaluating high-dimensional trial-level surrogate candidates across studies. The method offers several advantages over classical meta-analytic approaches in complex data settings. First, the nonparametric first stage avoids restrictive distributional assumptions on the individual participant data, enhancing robustness and applicability, particularly in small samples. The resulting within-study surrogacy measures are defined on a standardised scale and are invariant to monotone transformations, depending only on the relative ordering of observations within each study. This property is especially advantageous when integrating heterogeneous data sources measured on different scales (e.g., psychological assessments, multiplex assays, gene expression, or ELISPOT data). These standardised effect measures are directly comparable across markers and studies, providing interpretable and stable inputs for downstream meta-analysis. The evaluation stage further extends this framework by combining multiple markers into a composite signature, thereby strengthening surrogacy and reducing reliance on any single candidate in isolation. Finally, the use of an equivalence testing framework enables formal assessment of surrogacy against a clinically meaningful margin. While such approaches are well established in regulatory contexts (e.g., bioequivalence studies), they remain underutilised in surrogate endpoint evaluation \parencite{Saraf2014}.

We illustrated RISE-Meta on public vaccination data to evaluate gene expression biomarkers as trial-level surrogates for the clinically relevant downstream antibody response. Here, we identified a set of 7 genesets which acted as strong trial-level surrogates, both individually and in combination on unseen data. Ranking and visualising the top candidates from the marker-level meta-analysis (as in Figure \ref{fig:fig5}) allows further biological interpretation. The top of this list is enriched for interferon/antiviral (M127, M68, M75, M111.1), dendritic cell/antigen presentation (M165, M67, M95.1, S11, S5), and monocyte (M4.15, M11.0) modules, which refer to different components of the innate immune response which are expected to be observed in the first hours and days following vaccination \parencite{Hagan2022}. In addition, the interferon/antiviral interpretation reinforces the conclusions found in \textcite{Hughes2025}, where the RISE method (i.e. the first stage of RISE-Meta) was applied to a subset of a single trial (SDY1276). This application thus illustrates the proposed value of RISE-Meta for deriving strong, interpretable TLS in complex settings, where classical approaches cannot be applied without significant modification. 

RISE-Meta also has notable limitations. Firstly, it is known that the TOST procedure is conservative and underpowered \parencite{Meyners2012}, as we confirm in our simulation studies. Despite this, alternative procedures to the TOST are more conceptually complex and do not offer major power gains when applied to well-designed studies \parencite{Meyners2012}. Next, the choice of the equivalence margin $\varepsilon$ has long been discussed in equivalence testing \parencite{Walker2010}. A first strategy is simply to decide, based on the contextual costs and benefits of replacing the primary endpoint with a surrogate marker, a maximum bias one would tolerate when estimating the treatment effect on the former. Another, as exemplified in the application, is to use a data-driven approach based on the power within each study to detect a treatment effect on the primary endpoint based on the surrogate \parencite{Parast2024}. Alternatively, one may remove the dependence on p-values and equivalence margins entirely and consider the \textit{least equivalence allowable difference} (LEAD), which gives the smallest equivalence margin such that equivalence would be declared at a significance level $\alpha$ \parencite{Meyners2007}.

Several avenues remain open for the future development of RISE-Meta. First, Bayesian meta-analysis could be used to address the power limitation, especially when the number of studies is small \parencite{Roever2020}. Next, RISE-Meta uses a univariate screening procedure, but markers may belong to complex networks that only demonstrate surrogacy when considered simultaneously. A natural multivariate analogue of rank-based estimation and testing uses optimal transport, which could be adopted to extend the first stage of RISE-Meta \parencite{Huang2025}. Finally, although RISE-Meta is broadly applicable across continuous data, its utility could be widened by extending its first stage to other common data types, such as binary and survival data. 

\section*{Acknowledgements}\label{sec6}

This work is part of A.H.'s PhD thesis at the University of Bordeaux, co-supervised by B.P.H. and R.T., and supported by the University of Bordeaux's Digital Public Health Graduate School, funded by France's PIA 3 scheme (Investments for the Future – Project reference: 17-EURE-0019) through the Agence Nationale de la Recherche and University of Bordeaux’s France 2030 program. In addition, this work was supported by the Programme et Equipement Prioritaire de Recherche Santé Numérique (PEPR SN) project SMATCH (ref: 22-PESN-0003), the MUSICC (Mucosal Immunity in Human Coronavirus Challenge) project funded by the Coalition for Epidemic Preparedness Innovations (CEPI) and co-funded by the European Union’s Horizon Europe Programme, and in part by NIDDK grant R01DK118354 (Parast). We thank the Human Immunology Project Consortium for providing the ImmPort and ImmuneSpace platforms, as well as the original investigators and participants of the studies SDY56, SDY80, SDY180 and SDY1276. We disclose that generative AI tools were used to refine the phrasing of certain sections of this manuscript, and aid with the coding of the methods, but not to generate content.

\printbibliography[title={References}]

\section*{Supplementary Material}\label{sec7}

Web appendices and figures are provided in the Supplementary Material available on the publisher's website. All methods presented in this article are available from the \texttt{SurrogateRank} \texttt{R} package, available from \url{github.com/laylaparast/SurrogateRank}. A tutorial for RISE-Meta based on simulated data can be found at \url{www.laylaparast.com/surrogaterank}. The data that support the findings of this study are openly available from ImmPort and ImmuneSpace, with study reference numbers SDY56, SDY80, SDY180 and SDY1276. Instructions to download these data and code to reproduce all results in this manuscript are available at \url{github.com/arthurhughes27/RISE-meta} or on Zenodo at \url{10.5281/zenodo.20039496}. 

\end{refsection}

\newpage

\begin{figure}[htbp]
    \centering
    \includegraphics[width=\textwidth]{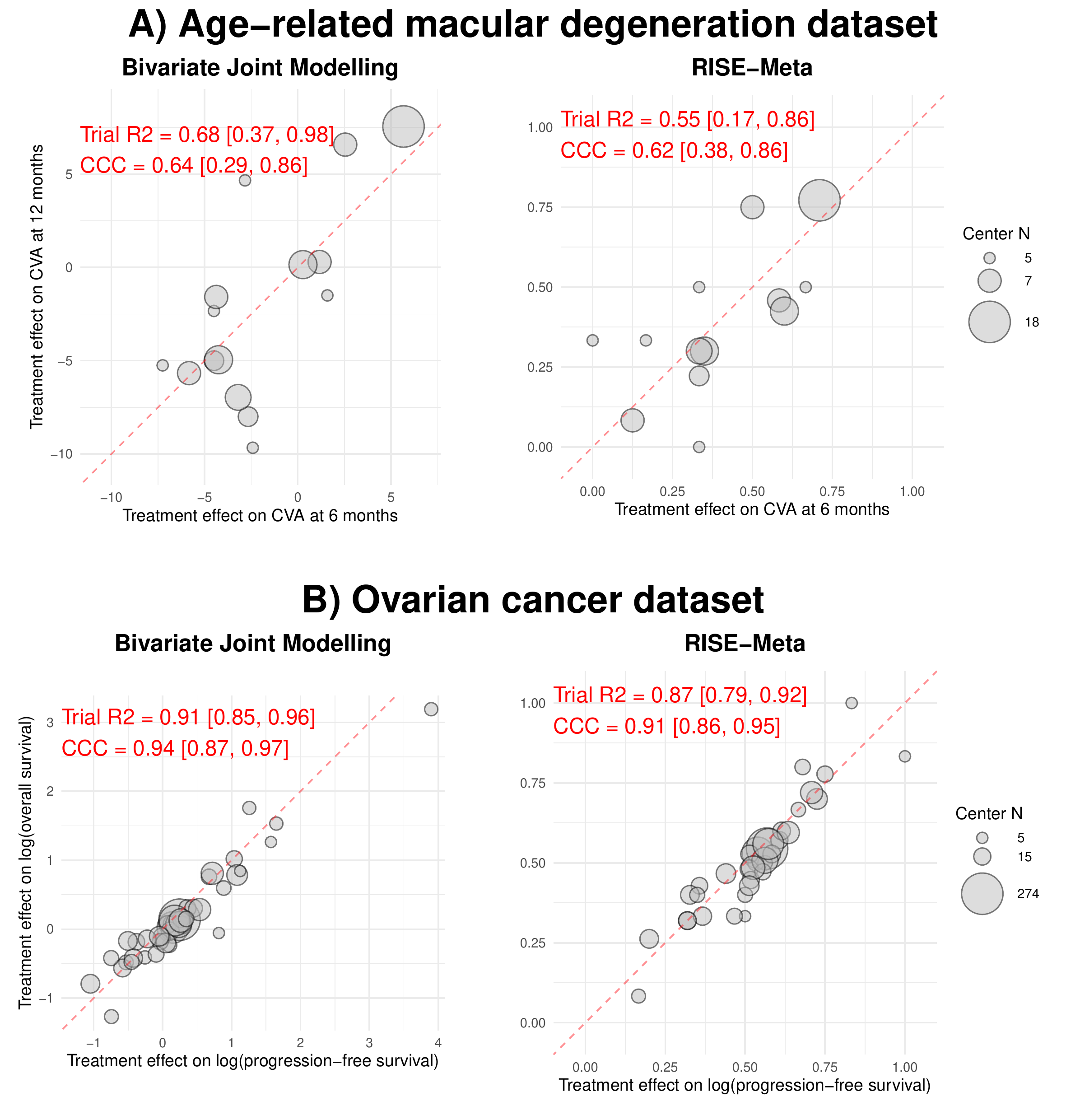}
    \caption{\textbf{Comparison of reference bivariate joint modelling approach to RISE-Meta applied to two low-dimensional datasets.} Panel \textbf{A)} gives the method comparison when applied to the \texttt{ARMD} dataset, whereas \textbf{B)} refers to the \texttt{Ovarian} dataset. Evaluation metrics and their 95\% CIs for each method and dataset are given in the top left corner. Overall, this demonstrates qualitative agreement between the two methods when applied to low-dimensional data. \textbf{Alt text:} Graphical comparison of the reference meta-analysis approach against RISE-Meta in two low-dimensional examples.}
    \label{fig:fig3}
\end{figure}

\begin{figure}[htbp]
    \centering
    \includegraphics[width=\textwidth]{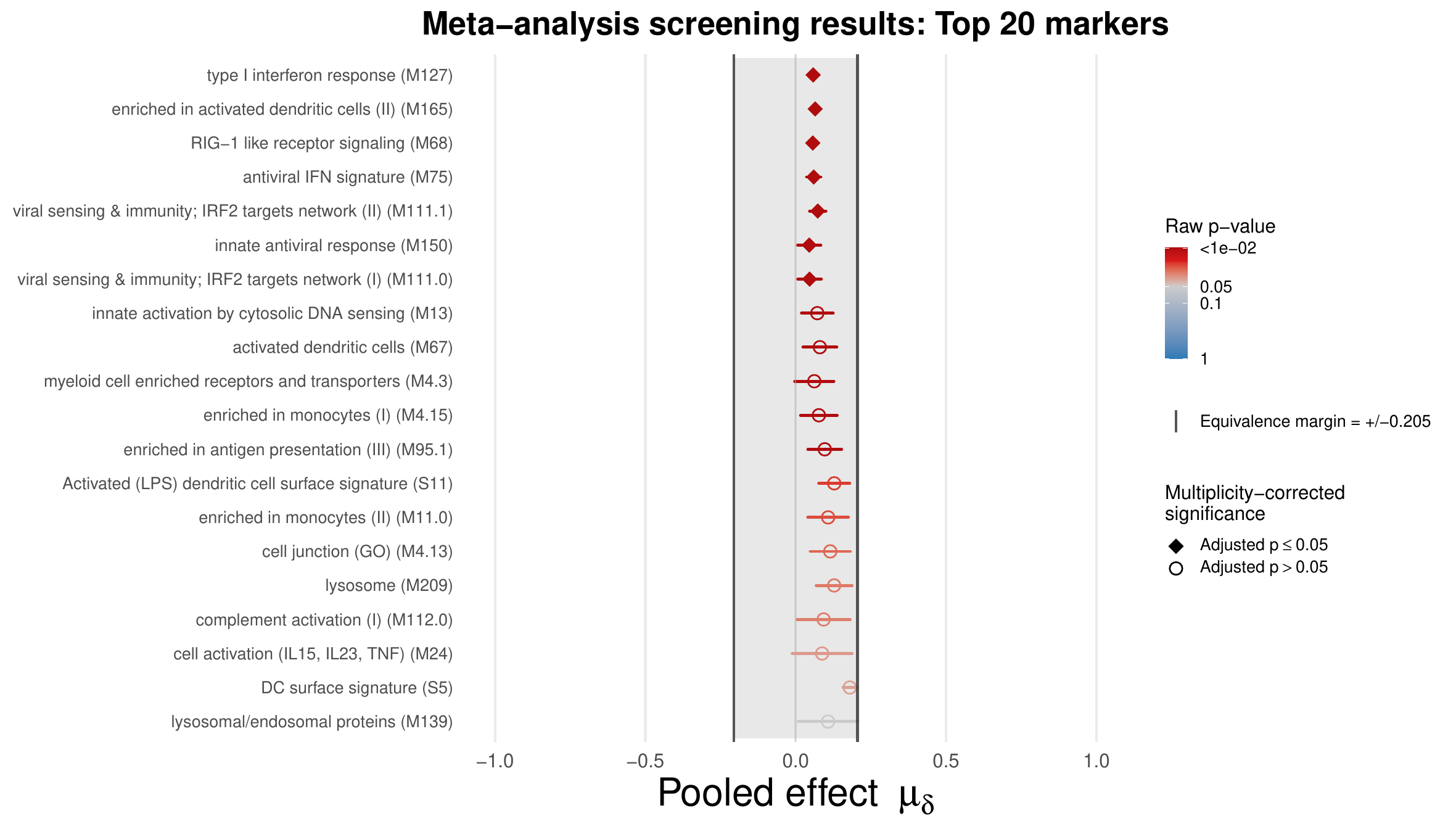}
    \caption{\textbf{Results from the marker-level meta-analysis.} This shows the pooled effect estimates for the top 20 markers by raw p-value. The shaded section is the equivalence region, delimited by the equivalence margins $\varepsilon \approx \pm 0.205$. The colour corresponds to the raw p-value, where shades of red indicate points whose $90\%$ CIs are contained entirely within these bounds. The shape of the point denotes the significance decision after adjusting for test multiplicity, which includes the top 7 markers. \textbf{Alt text:} Marker-level results of meta-analysis. }
    \label{fig:fig4}
\end{figure}

\begin{figure}[htbp]
    \centering
    \includegraphics[width=\textwidth]{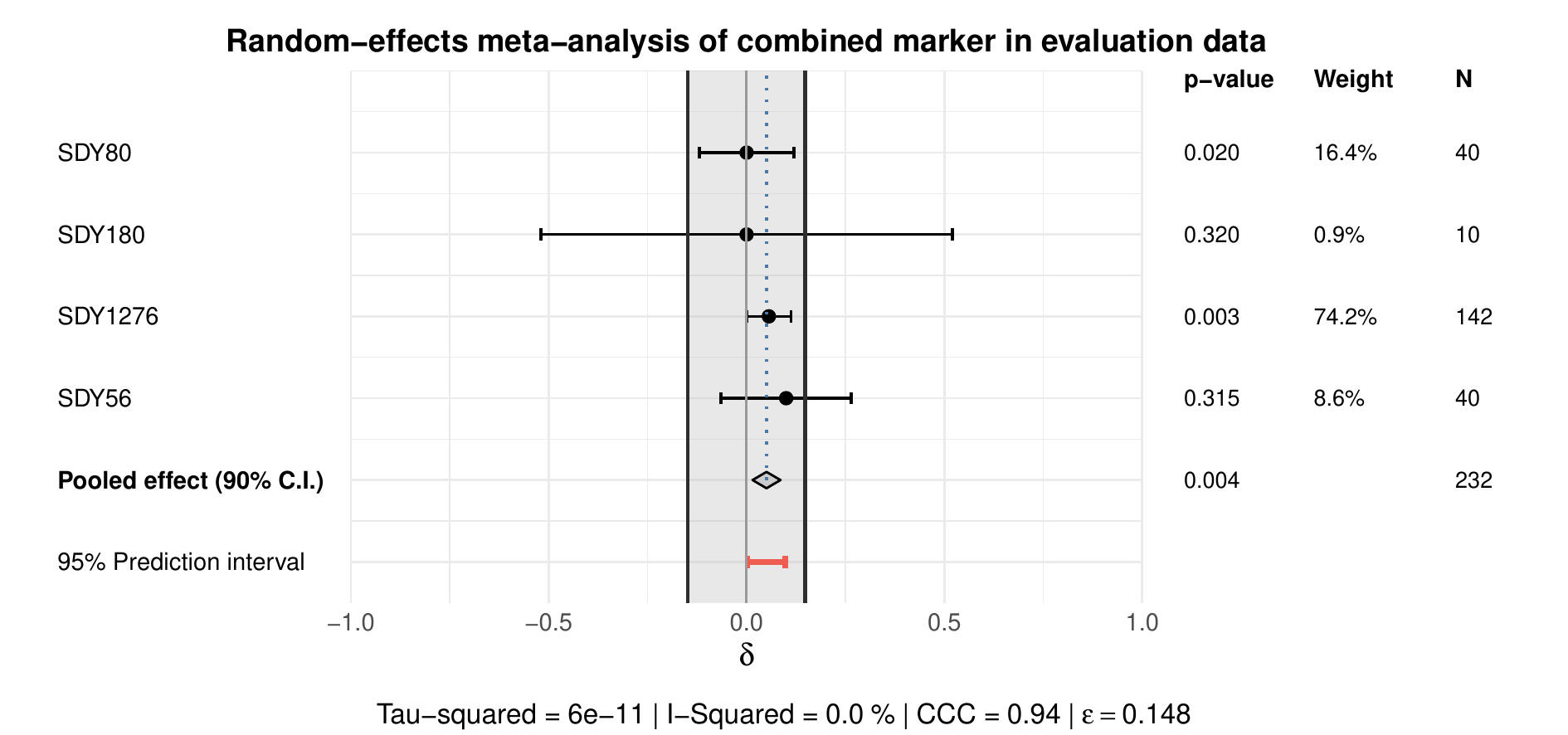}
    \caption{\textbf{Results from the evaluation of the 7-marker composite signature.} This shows the effect estimates $\delta$ for each of the 4 studies, and the pooled effect estimate. The within-study p-values, meta-analysis weights, and number of observations $N$ are given on the right hand side. The shaded section is the equivalence region, delimited by the equivalence margins $\varepsilon \approx \pm 0.148$. The pooled effect and its $90\%$ CI are contained with this region, corresponding to a TOST p-value of $\approx 0.004$. The bottom row gives the $95\%$ PI, which is also narrow due to the low between-study heterogeneity $\tau^{2} \approx 6e-11$. The CCC is 0.94, indicating strong concordance between the study-level effects on the primary endpoint and on the composite signature. \textbf{Alt text: } Meta-analysis for the combined 7-marker signature.}
    \label{fig:fig5}
\end{figure}

\clearpage

\begin{refsection}

\begin{center}
\textbf{\large Supplementary Material: Meta-Analysis of High-Dimensional Surrogate Markers}
\end{center}

\setcounter{figure}{0}
\setcounter{table}{0}
\setcounter{equation}{0}
\setcounter{section}{0}

\renewcommand{\figurename}{Web Figure}
\renewcommand{\tablename}{Web Table}

\renewcommand{\thefigure}{\arabic{figure}}
\renewcommand{\thetable}{\arabic{table}}
\renewcommand{\thesection}{S\arabic{section}}
\renewcommand{\theequation}{S\arabic{equation}}

\makeatother

\section{Web Appendix A: Methods}

\subsection{Notation}
For a reminder of the notation used throughout this Supplementary Material: we observe data from $m = 1,\dots,M$ trials, each with $i = 1,\dots,n_{m}$ individuals receiving the same binary treatment $A_{i,m} \in \{0,1\}$, where $a=1$ denotes active treatment and $a=0$ denotes control. Let $Y_{i,m}^{a}$ and $\boldsymbol{S}_{i,m}^{a} = (S_{i,1,m}^{a},\dots,S_{i,J,m}^{a})^{T}$ denote the primary endpoint and vector of $J$ common surrogate candidates, respectively, for individual $i$ in trial $m$ receiving treatment $a$. In the paired setting, for each individual and a given post-treatment timepoint, we have paired measurements $\{Y_{i,m}^{0},Y_{i,m}^{1},\boldsymbol{S}_{i,m}^{0},\boldsymbol{S}_{i,m}^{1}\}$, where the superscripts $a = 0$ and $a = 1$ refer to pre- and post-treatment measurements, respectively.

\subsection{Within-study surrogacy estimation}

Here, we discuss estimation of the within-study surrogacy parameter $\delta$, where again we omit marker and trial subscripts $j,m$ for notational clarity. We may obtain point estimates for these parameters using rank-based procedures. First, define a function $$G(x,y) = \begin{cases}
  1 , & \text{if } x> y,\\[2pt]
  \tfrac{1}{2}, & \text{if } x=y,\\[2pt]
  0, & \text{if } x<y.
\end{cases}$$

In the active treatment versus control context, we have $N^{1}, N^{0}$ treated and control individuals, respectively. Letting $i$ index treated and $j$ index control, we may estimate $$\widehat{U}_{Y} = \frac{1}{N^1 N^0} \sum_{i=1}^{N^1} \sum_{l=1}^{N^0} G(Y_{i}^{1}, Y_{l}^{0}), \qquad
\widehat{U}_{S} = \frac{1}{N^1 N^0} \sum_{i=1}^{N^1} \sum_{l=1}^{N^0} G(S_{i}^{1}, S_{l}^{0}),$$ 

In the paired data setting, we instead simply compare individuals post- and pre-treatment measurements, denoted with superscripts $1, 0$, respectively. If we have $N$ individuals, we may estimate $$\widehat{U}_{Y} = \frac{1}{N} \sum_{i=1}^{N} G(Y_{i}^{1}, Y_{i}^{0}), \qquad
\widehat{U}_{S} = \frac{1}{N} \sum_{i=1}^{N} G(S_{i}^{1}, S_{i}^{0}).$$

Detail of the estimation of the standard error in the active versus control setting is described in \textcite{Parast2024}, whereas its derivation in the paired data setting is found in the Supplementary Material of \textcite{Hughes2025}.

\subsection{Between-study variability estimation}

Here, we elaborate on the estimation of the between-study variability parameter $\tau^{2}$, where we again omit the marker subscript $j$ for clarity. The pooled mean effect depends on the between-study variance $\tau^2$ through the inverse-variance weights $$
w_m(\tau^2)=\frac{1}{\widehat{\sigma}_m^2+\tau^2},
\qquad
\mu(\tau^2)=\frac{\sum_{m=1}^M w_m(\tau^2)\widehat{\delta}_m}{\sum_{m=1}^M w_m(\tau^2)}.
$$
We estimate $\tau^2$ by restricted maximum likelihood (REML), as the nonnegative solution of the restricted score equation. Numerically, the score is evaluated by recomputing the generalised least-squares estimate $\mu(\tau^2)$ at each candidate value of $\tau^2$, and the root is sought on a bounded interval using a numerical root-finding algorithm with a small tolerance. If the score does not change sign on the search interval, or if root finding fails to converge, $\tau^2$ is instead obtained by direct bounded maximisation of the restricted log-likelihood over $\tau^2 \ge 0$. The resulting estimate is truncated at zero to enforce the parameter constraint.

\subsection{Conditions to avoid the surrogate paradox}

The surrogate paradox is a situation where the treatment effect on the surrogate is positive, the surrogate has positive association with the primary outcome, but the treatment effect on the primary outcome is negative. It is shown in \textcite{Parast2024}, that under the following conditions, we may avoid this:

\begin{align*}
(C1) \quad & P(Y^a > y \mid S^a = s) \text{ increases in } s \\
(C2) \quad & P(S^1 > s) \geq P(S^0 > s) \quad \forall s \\
(C3) \quad & P(Y^1 > y \mid S^1 = s) \geq P(Y^0 > y \mid S^0 = s) \quad \forall s
\end{align*}

(C1) states that the surrogate has a non-negative association with the outcome, (C2) that treatment has a non-negative effect on the surrogate, and (C3) that the residual treatment effect on the outcome after conditioning on the surrogate is non-negative. In particular, these conditions imply that $U_S > \frac{1}{2} \implies U_Y > \frac{1}{2}$, that is, that we can conclude a positive treatment effect on $Y$ based on a positive treatment effect observed on $S$. 

\subsection{Pooled effect and variance estimators}

The random-effects pooled effect is estimated as the inverse-variance weighted mean
$$
\widehat{\mu}
=
\frac{\sum_{m=1}^M \widehat{w}_m \widehat{\delta}_m}{\sum_{m=1}^M \widehat{w}_m},
\qquad
\widehat{w}_m=\frac{1}{\widehat{\sigma}_m^2+\widehat{\tau}^2},
$$
where $\widehat{\sigma}_m^2$ is the estimated within-study variance and $\widehat{\tau}^2$ is the estimated between-study variance. Under the conventional Wald-type approach, the variance of $\widehat{\mu}$ is estimated by
$$
\widehat{\mathrm{Var}}_{\mathrm{conv}}(\widehat{\mu})
=
\frac{1}{\sum_{m=1}^M \widehat{w}_m},
$$
leading to the normal-approximation confidence interval
$$
\widehat{\mu} \pm z_{1-\alpha/2}\sqrt{\widehat{\mathrm{Var}}_{\mathrm{conv}}(\widehat{\mu})}.
$$ where $z_{1-\alpha/2}$ denotes the $(1-\alpha/2)$-quantile of the standard normal distribution. 
This approach does not account for uncertainty in the estimation of $\tau^2$ and can therefore yield confidence intervals that are too narrow when the number of studies is small or heterogeneity is large. For this reason, we prefer the Hartung-Knapp-Sidik-Jonkman adjustment \parencite{Hartung2001}, which rescales the variance estimator by
$$
q=\frac{1}{M-1}\sum_{m=1}^M \widehat{w}_m(\widehat{\delta}_m-\widehat{\mu})^2,
\qquad
\widehat{\mathrm{Var}}_{\mathrm{HKSJ}}(\widehat{\mu})
=
q \cdot \frac{1}{\sum_{m=1}^M \widehat{w}_m}.
$$
Confidence intervals are then constructed using the $t$-distribution:
$$
\widehat{\mu} \pm t_{M-1,\,1-\alpha/2}\sqrt{\widehat{\mathrm{Var}}_{\mathrm{HKSJ}}(\widehat{\mu})}.
$$ where $t_{M-1,\,1-\alpha/2}$ denotes the $(1-\alpha/2)$-quantile of the Student's $t$-distribution with $M-1$ degrees of freedom.

\subsection{Concordance correlation coefficient}

Here we give further detail on the concordance correlation coefficient (CCC) \parencite{Lin1989} and its appropriateness in our setting. Here, the idea is to have a correlation-based evaluation metric which could be used as a complement to hypothesis testing and prediction intervals to summarise trial-level surrogacy. The classical methods of \textcite{Buyse2000} and \textcite{Molenberghs2002} use the coefficient of determination from a weighted linear regression of the trial-level effects on the primary endpoint and on the surrogate candidate, called $R^{2}_{trial}$, for this purpose, where $R^{2}_{trial} = 1$ corresponds to perfect surrogacy. 

In our framework, these treatment effects are defined on a common scale. This allows for a stricter notion of perfect surrogacy, namely that treatment effects are not only perfectly correlated across trials but also equal, i.e.\ $U_{Y,m} = U_{S,m}$ for all $m = 1,\dots,M$. Therefore, an appropriate metric of trial-level surrogacy should reflect both the correlation between trial-level effects and their agreement relative to the identity line. The CCC satisfies these criteria by jointly quantifying correlation and accuracy. Like standard correlation coefficients, it ranges in $\lbrack -1,1\rbrack $, with values near $1$ indicating strong concordance and values near $-1$ indicating strong discordance. The CCC is defined as $$ \mathrm{CCC} = 
\frac{2\,\mathrm{Cov}(\boldsymbol{U_Y}, \boldsymbol{U_S})}
     {\mathrm{Var}(\boldsymbol{U_Y}) + \mathrm{Var}(\boldsymbol{U_S}) + \left(\mathbb{E}[\boldsymbol{U_Y}] - \mathbb{E}[\boldsymbol{U_S}]\right)^2},
$$
where $\boldsymbol{U_{Y}} = (U_{Y,1},\dots,U_{Y,M})^{T}$ and $\boldsymbol{U_{S}} = (U_{S,1},\dots,U_{S,M})^{T}$ denote the vectors of trial-level treatment effects on the primary endpoint and generic surrogate marker $S$, respectively. In practice, $\mathrm{CCC}$ can be estimated using standard sample estimators of the moments across trials.

\subsection{Intraclass correlation coefficient}

Here, we briefly introduce the intraclass correlation coefficient (ICC), a reliability index derived from linear mixed-effects models, as an alternative validation metric to the CCC presented above. There are multiple forms of the ICC, defined by the model used to derive the index, the type of measurements, and the definition of agreement \parencite{Koo2016}. The estimation and interpretation of each differs, so it is important to choose an appropriate type based on the context.

In our case, we have paired measurements $(U_{Y,m},U_{S,m}) \quad \forall m \in \{1,\dots,M\}$. The appropriate form is therefore the two-way random-effects model, assessing absolute agreement between paired trial-level estimates at the individual level, which corresponds to the ICC(2,1) \parencite{Koo2016}. This may be estimated as follows.

$$
\bar{U_{S}} = \frac{1}{M}\sum_{m=1}^M U_{S,m}, \quad
\bar{U_{Y}} = \frac{1}{M}\sum_{m=1}^M U_{Y,m}, \quad
\bar{U} = \frac{1}{2M}\sum_{m=1}^M \left(U_{S,m} + U_{Y,m}\right),
$$

$$
\bar{U_{m}} = \frac{U_{S,m} + U_{Y,m}}{2}.
$$

$$
MS_R = \frac{2}{M-1} \sum_{m=1}^M \left(\bar{U_{m}} - \bar{U}\right)^2,
$$

$$
MS_C = M \left[ \left(\bar{U_{S}} - \bar{U}\right)^2 + \left(\bar{U_{Y}} - \bar{U}\right)^2 \right],
$$

$$
MS_E = \frac{1}{M-1} \sum_{m=1}^M \Big[
\left(U_{S,m} - \bar{U_{m}} - \bar{U_{S}} + \bar{U}\right)^2
+
\left(U_{Y,m} - \bar{U_{m}} - \bar{U_{Y}} + \bar{U}\right)^2
\Big],
$$

$$
\mathrm{ICC}(2,1)
=
\frac{MS_R - MS_E}
{MS_R + MS_E + \frac{2}{M}(MS_C - MS_E)}.
$$

The ICC(2,1) may be interpreted as the proportion of total variability in the paired trial-level estimates that is attributable to between-trial heterogeneity, rather than disagreement between the surrogate and primary treatment effect estimates. In other words, it quantifies the extent to which the surrogate and primary effects are consistent in capturing differences across trials, relative to their lack of agreement within trials.

We have a slight preference for the CCC as it provides a more direct and model-free measure of concordance with respect to the identity line $U_S = U_Y$, combining both correlation and systematic deviation. Nevertheless, the ICC(2,1) remains a valid complementary reliability-based summary of agreement.

\section{Web Appendix B: Simulation Study}

\subsection{Meta-analysis model specification and estimation}

Here, we investigate the impact of the meta-analysis model specification and estimation on the false positive rate control from the resulting hypothesis tests. In particular, we examine the choice to perform random-effects as opposed to fixed-effect meta-analysis. Separately, for random-effects meta-analysis, we evaluate the use of the HKSJ adjustments for the variance estimates of the pooled effects as opposed to the conventional normal approximation. 

Results are given in \cref{webfig4}. The fixed-effect model approach exceeds the nominal FPR for all but the smallest combinations of heterogeneity parameters. The random-effects model combined with the conventional approach to $\tau^{2}$ estimation controls the FPR relatively well, but exceeds it for low to moderate numbers of studies and heterogeneity. The random-effects model combined with the HKSJ approach controls the FPR well across all settings. 

Overall, these results support the use of the random-effects model combined with the HKSJ approach when the number of studies is low-to-moderate, and when there is any doubt about the assumptions underlying the fixed-effect model.

\subsection{Distribution of true mean}

As briefly explained in the main text, as our TOST equivalence testing procedure corresponds to testing a region of the parameter space, not all true parameter values will produce equal results. This is why, when investigating the test calibration, we generated invalid surrogates having true means at the boundaries of the valid region, corresponding to the least favourable configuration (LFC). It follows that, for invalid surrogates with means further into the invalid region of the parameter space, the empirical FPR will be at most as conservative as the same setting under the LFC. For the investigation of the power, we allowed the true means to vary uniformly across the valid region. 

In this section, we investigate the properties of the test under different settings of true mean generation. \cref{webfig5} presents the empirical FPR as a function of the true mean of invalid surrogate generation for 3 values of maximum between-study variance. As the true mean moves further into the invalid region, the empirical FPR quickly drops to 0. This is a consequence of the equivalence test setup, which must be calibrated to control the FPR in the LFC. 

Next, we investigated empirical power when fixing the true mean of valid surrogate generation. The results are presented in \cref{webfig6}. As expected, empirical power peaks when the true mean is generated at the greatest absolute distance from the equivalence region bounds i.e. $\mu = 0$, and goes to 0 as the mean approaches the bounds. The power improves for increasing numbers of studies. 

\subsection{Nonparametric simulation}

As noted in the main text, permutation-based simulations are useful for assessing test behaviour when the parametric model is uncertain and the data-generating distribution is unknown. The aim is to construct a permutation scheme on a real dataset that breaks the associations of interest while preserving key features of the observed data.

We used a subset of the Immune Signatures 2 dataset from the high-dimensional application in the main manuscript. This multi-study dataset contains paired pre- and post-influenza vaccination measurements. The primary endpoint was the mean antibody response across the three vaccine strains within each individual and timepoint. The surrogate candidates were the expression levels of $10{,}086$ genes, analysed at the gene-level rather than the geneset-level. The primary endpoint was measured at day 0 and day 28 ($\pm 7$ days), whereas the surrogate candidates were measured at day 0 and day 7. Although day 1 was used in the main application, day 7 was chosen here because it yielded a larger pool of available studies.

The permutation scheme was designed to break the association between the paired primary endpoint measurements and the paired surrogate measurements. Within each study, we permuted the paired primary endpoint values across individuals while leaving the surrogate pairs unchanged. This preserves study-level effects, within-individual correlation between pre- and post-vaccination measurements of the same variable, and between-marker correlation within individuals. Under this construction, the permuted datasets contain only true negatives, so this setting was used to assess false positive rate (FPR) control only.

Two design features were varied: the number of samples per study and the number of studies. These were controlled by subsampling from the permuted datasets. To study the effect of sample size:
\begin{enumerate}
    \item Define the sample size grid: $n^{*} \in \{5,10,\dots,25\}$.
    \item Restrict to studies with at least $25$ individuals: $M^{*} = \{m : n_m \ge 25\}$.
    \item For $b = 1,\dots,1000$:
    \begin{itemize}
        \item subsample $n^{*}$ individuals per study;
        \item apply the within-study permutation scheme;
        \item apply RISE-meta to obtain p-values $\{p_j^{(b)}\}_{j=1}^J$;
        \item compute the empirical FPR
        \[
        \widehat{\mathrm{FPR}}_b(n^{*}) = \frac{1}{J}\sum_{j=1}^J \mathbf{1}\!\left(p_j^{(b)} < \alpha\right).
        \]
    \end{itemize}
\end{enumerate}
A maximum sample size of $25$ was chosen so that the same set of studies remained available across the grid, leaving $6$ studies.

To study the effect of the number of studies:
\begin{enumerate}
    \item Restrict to studies with at least $15$ individuals: $M^{*} = \{m : n_m \ge 15\}$.
    \item Define the study grid: $m^{*} \in \{2,\dots,10\}$.
    \item For $b = 1,\dots,1000$:
    \begin{itemize}
        \item sample a subset of studies $S_b \subset M^{*}$ with $|S_b| = m^{*}$;
        \item subsample $15$ individuals per study;
        \item apply the within-study permutation scheme;
        \item apply RISE-meta to obtain p-values $\{p_j^{(b)}\}_{j=1}^J$;
        \item compute the empirical FPR
        \[
        \widehat{\mathrm{FPR}}_b(m^{*}) = \frac{1}{J}\sum_{j=1}^J \mathbf{1}\!\left(p_j^{(b)} < \alpha\right).
        \]
    \end{itemize}
\end{enumerate}
Here, $15$ individuals per study was chosen so that up to $10$ studies were available.

\cref{webfig8} shows the results for the sample-size investigation. The average empirical FPR remained well below the nominal level $\alpha = 0.05$ across all sample sizes, and decreased towards $0$ as the number of observations increased. This conservatism is expected: unlike in the parametric simulations, the invalid surrogates in this real-data setting are unlikely to lie at the least favourable configuration used in the parametric case, and their true mean is probably farther into the invalid region (see \cref{webfig5} for a demonstration).

\cref{webfig7} shows the results for the study-number investigation. False positive rate control was maintained across all values of $m^{*}$. For the smallest number of studies, the empirical FPR occasionally exceeded the nominal level, with a maximum of $0.10$ when only $2$ studies were used. As the number of studies increased, the empirical mean again approached $0$.

Overall, these results support the statistical validity of the equivalence testing in RISE-Meta in realistic scenarios under unknown data generating processes. 

\section{Web Appendix C: Data applications}

\subsection{Low-dimensional examples}

\subsubsection{Description of bivariate joint modelling approach}

 \textcite{Buyse2000} and \textcite{Molenberghs2002} describe an approach using bivariate mixed-effects models to assess trial-level surrogacy in the meta-analytic framework. Using our notation for an arbitrary surrogate candidate $S$ for individual $i$ in trial $m$, they consider models $$S_{i,m} = \mu_{S} + m_{S,m} + (\alpha + a_{m})A_{i,m} + \epsilon_{S_{i,m}},$$
$$Y_{i,m} = \mu_{Y} + m_{Y,m} + (\beta + b_{m})A_{i,m} + \epsilon_{Y_{i,m}}$$ where $\mu_{S}, \mu_{Y}$ are fixed intercepts, $m_{S,m}, m_{Y,m}$ are trial-specific random intercepts, $\alpha, \beta$ are fixed treatment effects, and $a_{m}, b_{m}$ are trial-specific random treatment effects. The random-effects $\boldsymbol{V} = (m_{S,m}, m_{Y,m}, a_{m}, b_{m})^{T}$ are assumed to be multivariate normally distributed $V \sim \mathcal{N}(\boldsymbol{0}, D)$, where $$D = \begin{pmatrix}
d_{SS} & d_{SY} & d_{Sa} & d_{Sb} \\
d_{SY} & d_{YY} & d_{Ya} & d_{Yb} \\
d_{Sa} & d_{Ya} & d_{aa} & d_{ab} \\
d_{Sb} & d_{Yb} & d_{ab} & d_{bb}
\end{pmatrix} $$

and the error terms $\boldsymbol{\epsilon}_{i,m} = (\epsilon_{S_{i,m}}, \epsilon_{Y_{i,m}})^{T}$ are bivariate normally distributed $\boldsymbol{\epsilon}_{i,m} \sim \mathcal{N}(\boldsymbol{0}, \Sigma)$ where $$\Sigma = \begin{pmatrix}
\sigma_{SS} & \sigma_{SY}  \\
\sigma_{SY} & \sigma_{YY} 
\end{pmatrix}$$

This gives the trial-level surrogacy metric $$R_{trial}^{2} = \dfrac{\begin{pmatrix} d_{Sb} \\ d_{ab} \end{pmatrix}^{T}  \begin{pmatrix} d_{SS} & d_{Sa} \\ d_{Sa} & d_{aa} \end{pmatrix}^{-1}  \begin{pmatrix} d_{Sb} \\ d_{ab} \end{pmatrix} }{d_{bb}}$$

However, as described in \textcite{Tibaldi2003, 2005}, this approach often results in computational problems. Indeed, in our two data examples from the main manuscript, we were unable to fit these models using the \texttt{BimixedContCont()} function from the \texttt{Surrogate} package, either as described above or in a form with reduced complexity of the random-effects structure. This was due to numerical convergence issues arising from the estimation of $D$ as a singular covariance matrix. An alternative approach is to instead fit a simplified modelling strategy using a two-stage bivariate fixed-effect modelling approach, implemented through the \texttt{BifixedContCont()} function. 

In the first stage, this uses bivariate fixed-effect model $$S_{i,m} = \mu_{S,m} + \alpha_{m} A_{i,m} + \epsilon_{S_{i,m}},$$
$$Y_{i,m} = \mu_{Y,m} + \beta_{m} A_{i,m} + \epsilon_{Y_{i,m}}$$ where $\mu_{S,m}, \mu_{Y,m}$ are trial-specific fixed intercepts, $\alpha_{m}, \beta_{m}$ are trial-specific fixed treatment effects, and the error terms are distributed as above.

In the second stage, the trial-specific treatment effects on the primary endpoint are modelled as a function of the trial-specific surrogate intercepts and treatment effects, estimated from the first stage, as $$\widehat{\beta_{m}} = \kappa_{0} + \kappa_{1}\widehat{\mu}_{S,m} + \kappa_{2}\widehat{\alpha}_{m} + \epsilon_{i}$$

This model is fit using weighted least squares, where the weights are the within-study sample sizes. Then, the $R_{trial}^{2}$ is taken as the coefficient of determination from this weighted linear regression fit. A 95$\%$ confidence interval around this quantity can be estimated using the delta method. 

\subsubsection{Discussion on analysis choices}

A few choices in the low-dimensional data analysis examples merit discussion. We note that we are using these data sets as illustrations for benchmarking the qualitative conclusions of the two methods, and the resulting analysis should be interpreted with caution. In addition, in these analyses, our analysis choices follow the lead of previous works which have already used these datasets (or sub/supersets thereof) \parencite{Buyse2000, Molenberghs2002, Tibaldi2003, 2005}. 

Firstly, we discuss the fact that we choose to use the treatment center as the unit of analysis for the analysis of the data. This choice enables application of the meta-analytic frameworks, which require between-unit variability. In the case of the \texttt{ARMD} data, the data arise from a single RCT, and therefore cannot be analysed with meta-analytical methods without further choosing a smaller investigative unit. The \texttt{Ovarian} dataset comes from 4 RCTs, but information on these is not available in the dataset accessible from the \texttt{Surrogate} package. Further discussion on the choice of units for meta-analytical evaluation of surrogate markers can be found in \textcite{Abrahantes2004}. 

Another point of discussion is the treatment of survival times as continuous outcomes. Here, we used log transformations on the survival times to render them continuous and unbounded. However, we note that an approach tailored to survival data, and in particular taking into account censoring, is a more appropriate way to analyse these data.

\subsubsection{Evaluation metrics and confidence intervals}

We estimate a $R^{2}_{trial}$ statistic for the RISE-Meta results in a similar spirit to the bivariate fixed-effect modelling approach, in two stages. First, we compute estimates $\boldsymbol{\widehat{U}_{Y}} = (\widehat{U}_{Y,1},\dots,\widehat{U}_{Y,M})^{T}$ and $\boldsymbol{\widehat{U}_{S}} = (\widehat{U}_{S,1},\dots,\widehat{U}_{S,M})^{T}$ and then fit a model $$\widehat{U}_{Y,m} = \kappa_{0} + \kappa_{1}\widehat{U}_{S,M} + \epsilon_{i}$$ with weighted least squares, where the weights are the sample sizes $n_{m}$, taking the $R^{2}$ statistic from this fit.

In the case of the bivariate fixed-effects model approach and the $R^{2}_{trial}$ statistic, the $95\%$ CI could be extracted directly from the \texttt{BifixedContCont()} function, and is estimated via the delta method. Otherwise, the variance of the $R^{2}_{trial}$ for RISE-Meta and the CCC for both methods is estimated with bias-corrected and accelerated bootstrapping \parencite{Efron1987}. Specifically, we generate $B = 2000$ datasets by re-sampling with replacement $M$ pairs $(U_{Y,m}, U_{S,m})$. 

\subsection{High-dimensional example}

In this section, we discuss the high-dimensional data application example, including a fuller description of the dataset, as well as sensitivity and supplementary analyses. 

\subsubsection{Description of the Immune Signatures 2 dataset and preprocessing}

\begin{figure}[ht]
\centerline{\includegraphics[height = 11cm]{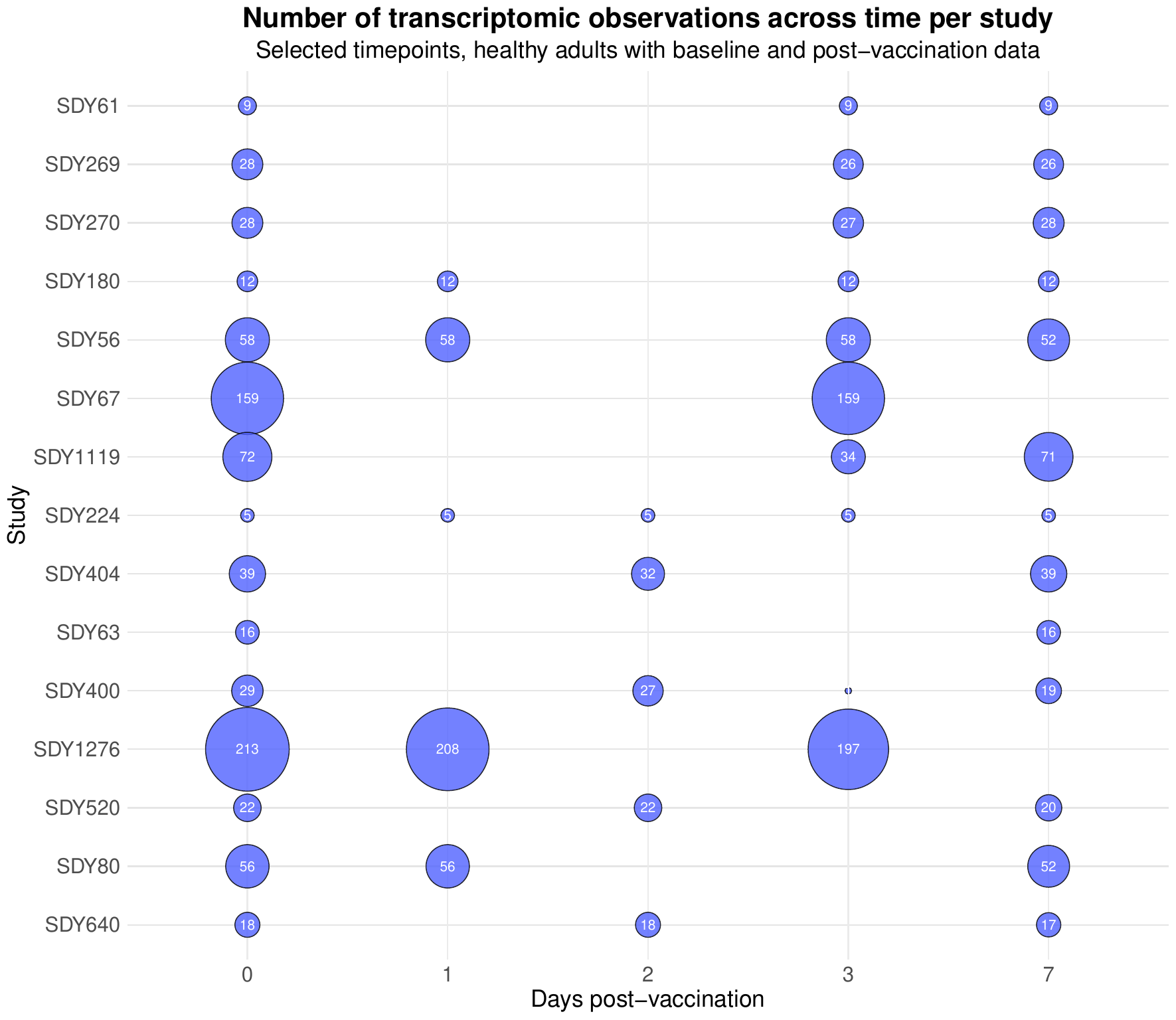}}
\caption{\textbf{Available timepoints and sample sizes per inactivated influenza study in the IS2 data}. The sample sizes denote the unique individuals with available immune response data (nAb or HAI) at each timepoint. \label{webfig1}}
\end{figure}

\begin{figure}[ht]
\centerline{\includegraphics[height = 11cm]{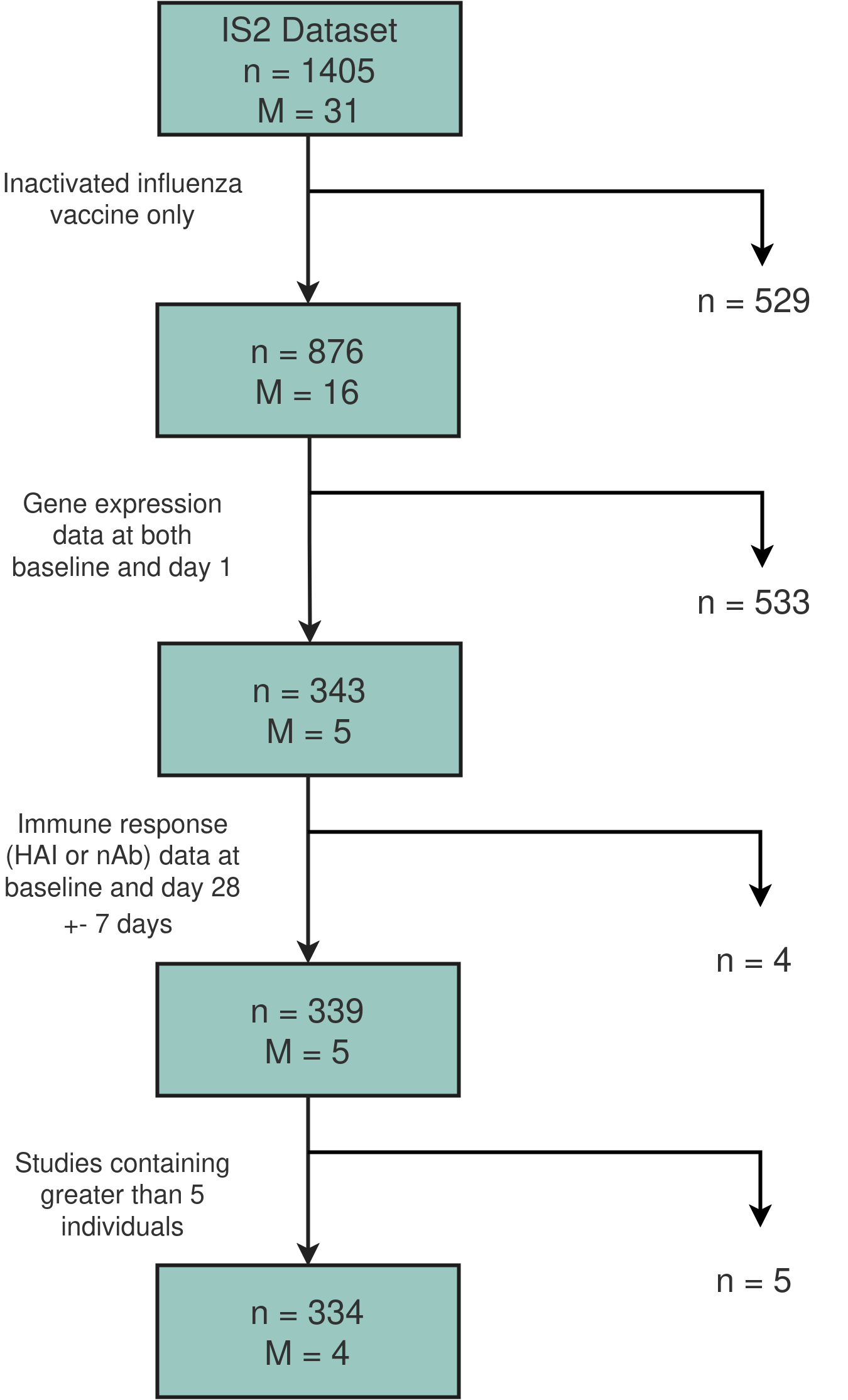}}
\caption{\textbf{Flowchart showing the filtration of the IS2 dataset prior to analysis}. $n$ denotes the total number of unique individuals and $M$ the number of studies in the dataset at each stage.\label{webfig2}}
\end{figure}

Here we give a fuller description of the subset of the Immune Signatures 2 (IS2) dataset considered as an illustrative example for the application of RISE-Meta. The Immune Signatures 2 dataset is a release of the Immune Signatures Data Resource (ISDR), which is a standardised compendium of systems vaccinology datasets. A description of the full dataset can be found in \textcite{DirayArce2022}. For our purposes, we considered only datasets on inactivated influenza vaccination, since this vaccine had available data from multiple studies, necessary for meta-analysis. All studies consisted of longitudinal measurements of gene expression markers. Therefore, it was necessary to make a choice of the post-vaccination timepoint used for the analysis. A selection of the candidate timepoints available for each of the trivalent inactivated influenza studies, and the valid samples at each, is presented in \cref{webfig1}. Day 1 was chosen as the post-treatment timepoint of interest due to our \textit{a priori} knowledge and experience with these datasets suggesting the most promising signal would likely be found soon after vaccination. A flowchart showing the preprocessing steps taken to filter the individuals from the IS2 dataset is given in \cref{webfig2}. This includes the removal of individuals without gene expression measurements at both baseline and day 1, immune response measurements at both baseline and day 28 $\pm 7$ days, and studies containing 5 or fewer individuals. The latter filtering stage is important as, below a certain number of individuals, the RISE method cannot estimate the sampling variability of the surrogacy parameter $\delta$. 

\subsubsection{Geneset pre-processing}

Initially, the BTMs consisted of $J = 346$ genesets. For better interpretability, we preprocessed the BTM genesets to contain only modules with informative titles, which involved the removal of any geneset titled ``TBA" (to be attributed). In addition, one geneset that had no member genes in common with the $10,086$ initial input features was removed. After these filtering steps, we obtained $J = 258$ features as inputs into RISE-Meta. 

\subsubsection{Fixed-effect analysis}

Here, we perform the same analysis as in the main manuscript, but instead do fixed-effect meta-analysis. \cref{webfig9} shows the results of the marker-level meta-analysis. The standard errors for each marker are smaller compared to the random-effects meta-analysis, leading to more significant markers (16 compared to 7). \cref{webfig10} shows the results of the evaluation of the 16-marker signature on the $33\%$ of unseen data. Interestingly, the point estimates are less consistent using this 16-marker signature compared to the 7-marker signature identified in the RE analysis. In addition, the CI is wider and the CCC is smaller. This is possibly a reflection of the additional noise incorporated into the signature due to the more relaxed significance criteria when using the FE method. Overall, however, the qualitative conclusions of the markers forming this signature stay the same as in the RE analysis, with this signature being dominated by antiviral, dendritic cell, and monocyte signals.

\subsubsection{Impact of geneset choice and aggregation method}

Here, we perform supplementary analyses to evaluate the qualitative impact of the choice of geneset definition and aggregation method. In particular, we investigate the use of a different set of genesets, called the BloodGen3Modules (BG3M), as presented in \textcite{Rinchai2021}. The BG3Ms were processed in a similar way to the BTMs, where an initial number of $J = 382$ genesets was filtered by removing sets with uninformative titles (i.e. genesets titled "TBD"), resulting in $J = 205$ inputs to RISE-Meta. In this analysis, we continue to use the mean as the aggregation function. 

\cref{webfig11} presents the results of the marker-level meta-analysis when using the BG3M genesets. Here, only 1 geneset is significant after multiplicity correction. Qualitatively, the top of this list is still dominated by interferon signals, with several other innate immune modules further down (monocytes, inflammation). \cref{webfig12} shows the evaluation of this single marker on unseen data, where the null hypothesis for non-equivalence of the pooled effect is rejected. 

In addition, we performed RISE-Meta when choosing not to aggregate at all to the geneset-level. \cref{webfig13} shows the top results for the marker-level meta-analysis in this analysis. Although many markers have pooled effects well within the equivalence margin, due to the severe multiplicity penalty after performing over 10,000 hypothesis tests, no marker is significant after adjustment. This points to an advantage of the geneset approach, which allows for a prior dimension reduction of the candidate biomarkers. 

\printbibliography[title={Supplementary Material References}]

@Article{Hughes2025,
  author     = {Hughes, Arthur and Parast, Layla and Thiébaut, Rodolphe and Hejblum, Boris P.},
  journal    = {Statistics in Medicine},
  title      = {RISE: Two‐Stage Rank‐Based Identification of High‐Dimensional Surrogate Markers Applied to Vaccinology},
  year       = {2025},
  issn       = {1097-0258},
  month      = sep,
  number     = {20–22},
  volume     = {44},
  doi        = {10.1002/sim.70241},
  publisher  = {Wiley},
  ranking    = {rank5},
  readstatus = {read},
}

@Article{Parast2024,
  author     = {Parast, Layla and Cai, Tianxi and Tian, Lu},
  journal    = {Biometrics},
  title      = {A rank-based approach to evaluate a surrogate marker in a small sample setting},
  year       = {2024},
  issn       = {1541-0420},
  month      = jan,
  number     = {1},
  volume     = {80},
  doi        = {10.1093/biomtc/ujad035},
  publisher  = {Oxford University Press (OUP)},
  ranking    = {rank5},
  readstatus = {read},
}

@Article{Hartung2001,
  author     = {Hartung, Joachim and Knapp, Guido},
  journal    = {Statistics in Medicine},
  title      = {A refined method for the meta‐analysis of controlled clinical trials with binary outcome},
  year       = {2001},
  issn       = {1097-0258},
  month      = dec,
  number     = {24},
  pages      = {3875--3889},
  volume     = {20},
  doi        = {10.1002/sim.1009},
  publisher  = {Wiley},
  ranking    = {rank3},
  readstatus = {read},
}

@Article{Lin1989,
  author     = {Lin, Lawrence I-Kuei},
  journal    = {Biometrics},
  title      = {A Concordance Correlation Coefficient to Evaluate Reproducibility},
  year       = {1989},
  issn       = {0006-341X},
  month      = mar,
  number     = {1},
  pages      = {255},
  volume     = {45},
  doi        = {10.2307/2532051},
  publisher  = {JSTOR},
  ranking    = {rank4},
  readstatus = {read},
}

@Article{Buyse2000,
  author    = {Buyse, Marc and Molenberghs, Geert and Burzykowski, Tomasz and Renard, Didier and Geys, Helena},
  journal   = {Drug Information Journal},
  title     = {Statistical Validation of Surrogate Endpoints: Problems and Proposals},
  year      = {2000},
  issn      = {2164-9200},
  month     = apr,
  number    = {2},
  pages     = {447--454},
  volume    = {34},
  doi       = {10.1177/009286150003400213},
  publisher = {Springer Science and Business Media LLC},
}

@Article{IntHout2014,
  author    = {IntHout, Joanna and Ioannidis, John PA and Borm, George F},
  journal   = {BMC Medical Research Methodology},
  title     = {The Hartung-Knapp-Sidik-Jonkman method for random effects meta-analysis is straightforward and considerably outperforms the standard DerSimonian-Laird method},
  year      = {2014},
  issn      = {1471-2288},
  month     = feb,
  number    = {1},
  volume    = {14},
  doi       = {10.1186/1471-2288-14-25},
  publisher = {Springer Science and Business Media LLC},
}

@Article{IntHout2016,
  author     = {IntHout, Joanna and Ioannidis, John P A and Rovers, Maroeska M and Goeman, Jelle J},
  journal    = {BMJ Open},
  title      = {Plea for routinely presenting prediction intervals in meta-analysis},
  year       = {2016},
  issn       = {2044-6055},
  month      = jul,
  number     = {7},
  pages      = {e010247},
  volume     = {6},
  doi        = {10.1136/bmjopen-2015-010247},
  publisher  = {BMJ},
  ranking    = {rank5},
  readstatus = {read},
}

@Article{Molenberghs2002,
  author    = {Molenberghs, Geert and Buyse, Marc and Geys, Helena and Renard, Didier and Burzykowski, Tomasz and Alonso, Ariel},
  journal   = {Controlled Clinical Trials},
  title     = {Statistical challenges in the evaluation of surrogate endpoints in randomized trials},
  year      = {2002},
  issn      = {0197-2456},
  month     = dec,
  number    = {6},
  pages     = {607--625},
  volume    = {23},
  doi       = {10.1016/s0197-2456(02)00236-2},
  publisher = {Elsevier BV},
}

@Article{Benjamini1995,
  author    = {Benjamini, Yoav and Hochberg, Yosef},
  journal   = {Journal of the Royal Statistical Society Series B: Statistical Methodology},
  title     = {Controlling the False Discovery Rate: A Practical and Powerful Approach to Multiple Testing},
  year      = {1995},
  issn      = {1467-9868},
  month     = jan,
  number    = {1},
  pages     = {289--300},
  volume    = {57},
  doi       = {10.1111/j.2517-6161.1995.tb02031.x},
  publisher = {Oxford University Press (OUP)},
}

@Article{DirayArce2022,
  author    = {Diray-Arce, Joann and Miller, Helen E. R. and Henrich, Evan and Gerritsen, Bram and Mulè, Matthew P. and Fourati, Slim and Gygi, Jeremy and Hagan, Thomas and Tomalin, Lewis and Rychkov, Dmitry and Kazmin, Dmitri and Chawla, Daniel G. and Meng, Hailong and Dunn, Patrick and Campbell, John and Deckhut-Augustine, Alison and Gottardo, Raphael and Haddad, Elias K. and Hafler, David A. and Harris, Eva and Farber, Donna and Levy, Ofer and McElrath, Julie and Montgomery, Ruth R. and Peters, Bjoern and Rahman, Adeeb and Reed, Elaine F. and Rouphael, Nadine and Fernandez-Sesma, Ana and Sette, Alessandro and Stuart, Ken and Togias, Alkis and Tsang, John S. and Sarwal, Minnie and Tsang, John S. and Levy, Ofer and Pulendran, Bali and Sekaly, Rafick and Floratos, Aris and Gottardo, Raphael and Kleinstein, Steven H. and Suárez-Fariñas, Mayte},
  journal   = {Scientific Data},
  title     = {The Immune Signatures data resource, a compendium of systems vaccinology datasets},
  year      = {2022},
  issn      = {2052-4463},
  month     = oct,
  number    = {1},
  volume    = {9},
  doi       = {10.1038/s41597-022-01714-7},
  publisher = {Springer Science and Business Media LLC},
}

@Article{Li2013,
  author    = {Li, Shuzhao and Rouphael, Nadine and Duraisingham, Sai and Romero-Steiner, Sandra and Presnell, Scott and Davis, Carl and Schmidt, Daniel S and Johnson, Scott E and Milton, Andrea and Rajam, Gowrisankar and Kasturi, Sudhir and Carlone, George M and Quinn, Charlie and Chaussabel, Damien and Palucka, A Karolina and Mulligan, Mark J and Ahmed, Rafi and Stephens, David S and Nakaya, Helder I and Pulendran, Bali},
  journal   = {Nature Immunology},
  title     = {Molecular signatures of antibody responses derived from a systems biology study of five human vaccines},
  year      = {2013},
  issn      = {1529-2916},
  month     = dec,
  number    = {2},
  pages     = {195--204},
  volume    = {15},
  doi       = {10.1038/ni.2789},
  publisher = {Springer Science and Business Media LLC},
}

@Article{Ball2020,
  author    = {Ball, Tali M. and Squeglia, Lindsay M. and Tapert, Susan F. and Paulus, Martin P.},
  journal   = {Biological Psychiatry: Cognitive Neuroscience and Neuroimaging},
  title     = {Double Dipping in Machine Learning: Problems and Solutions},
  year      = {2020},
  issn      = {2451-9022},
  month     = mar,
  number    = {3},
  pages     = {261--263},
  volume    = {5},
  doi       = {10.1016/j.bpsc.2019.09.003},
  publisher = {Elsevier BV},
}

@Article{Schuirmann1987,
  author    = {Schuirmann, Donald J.},
  journal   = {Journal of Pharmacokinetics and Biopharmaceutics},
  title     = {A comparison of the Two One-Sided Tests Procedure and the Power Approach for assessing the equivalence of average bioavailability},
  year      = {1987},
  issn      = {0090-466X},
  month     = dec,
  number    = {6},
  pages     = {657--680},
  volume    = {15},
  doi       = {10.1007/bf01068419},
  publisher = {Springer Science and Business Media LLC},
}

@Article{Tseng2012,
  author    = {Tseng, George C. and Ghosh, Debashis and Feingold, Eleanor},
  journal   = {Nucleic Acids Research},
  title     = {Comprehensive literature review and statistical considerations for microarray meta-analysis},
  year      = {2012},
  issn      = {0305-1048},
  month     = jan,
  number    = {9},
  pages     = {3785--3799},
  volume    = {40},
  doi       = {10.1093/nar/gkr1265},
  publisher = {Oxford University Press (OUP)},
}

@Article{Choi2003,
  author    = {Choi, Jung Kyoon and Yu, Ungsik and Kim, Sangsoo and Yoo, Ook Joon},
  journal   = {Bioinformatics},
  title     = {Combining multiple microarray studies and modeling interstudy variation},
  year      = {2003},
  issn      = {1367-4803},
  month     = jul,
  number    = {suppl},
  pages     = {i84--i90},
  volume    = {19},
  doi       = {10.1093/bioinformatics/btg1010},
  publisher = {Oxford University Press (OUP)},
}

@Article{Agniel2022,
  author    = {Agniel, Denis and Hejblum, Boris P and Thiébaut, Rodolphe and Parast, Layla},
  journal   = {Biostatistics},
  title     = {Doubly robust evaluation of high-dimensional surrogate markers},
  year      = {2022},
  issn      = {1468-4357},
  month     = jul,
  number    = {4},
  pages     = {985--999},
  volume    = {24},
  doi       = {10.1093/biostatistics/kxac020},
  publisher = {Oxford University Press (OUP)},
}

@Article{Zhou2022,
  author    = {Zhou, Ruixuan Rachel and Zhao, Sihai Dave and Parast, Layla},
  journal   = {Statistics in Medicine},
  title     = {Estimation of the proportion of treatment effect explained by a high‐dimensional surrogate},
  year      = {2022},
  issn      = {1097-0258},
  month     = feb,
  number    = {12},
  pages     = {2227--2246},
  volume    = {41},
  doi       = {10.1002/sim.9352},
  publisher = {Wiley},
}

@Article{Saraf2014,
  author    = {Saraf, Sanatan and Mathew, Thomas and Roy, Anindya},
  journal   = {Journal of Biopharmaceutical Statistics},
  title     = {Statistical Validation of Surrogate Endpoints: Another Look at the Prentice Criterion and Other Criteria},
  year      = {2014},
  issn      = {1520-5711},
  month     = nov,
  number    = {6},
  pages     = {1234--1246},
  volume    = {25},
  doi       = {10.1080/10543406.2014.971174},
  publisher = {Informa UK Limited},
}

@Misc{Huang2025,
  author    = {Huang, Zhen and Sen, Bodhisattva},
  title     = {Multivariate Distribution-Free Nonparametric Testing: Generalizing Wilcoxon's Tests via Optimal Transport},
  year      = {2025},
  copyright = {Creative Commons Attribution 4.0 International},
  doi       = {10.48550/ARXIV.2503.12236},
  publisher = {arXiv},
}

@Article{Daniels1997,
  author    = {Daniels, Michael J. and Hughes, Michael D.},
  journal   = {Statistics in Medicine},
  title     = {Meta-analysis for the evaluation of potential surrogate markers},
  year      = {1997},
  issn      = {1097-0258},
  month     = sep,
  number    = {17},
  pages     = {1965--1982},
  volume    = {16},
  doi       = {10.1002/(sici)1097-0258(19970915)16:17<1965::aid-sim630>3.0.co;2-m},
  publisher = {Wiley},
}

@Article{Alonso2004,
  author    = {Alonso, Ariel and Geys, Helena and Molenberghs, Geert and Kenward, Michael G. and Vangeneugden, Tony},
  journal   = {Biometrics},
  title     = {Validation of Surrogate Markers in Multiple Randomized Clinical Trials with Repeated Measurements: Canonical Correlation Approach},
  year      = {2004},
  issn      = {1541-0420},
  month     = dec,
  number    = {4},
  pages     = {845--853},
  volume    = {60},
  doi       = {10.1111/j.0006-341x.2004.00239.x},
  publisher = {Oxford University Press (OUP)},
}

@Article{Korn2016,
  author    = {Korn, E.L. and Sachs, M.C. and McShane, L.M.},
  journal   = {Annals of Oncology},
  title     = {Statistical controversies in clinical research: assessing pathologic complete response as a trial-level surrogate end point for early-stage breast cancer},
  year      = {2016},
  issn      = {0923-7534},
  month     = jan,
  number    = {1},
  pages     = {10--15},
  volume    = {27},
  doi       = {10.1093/annonc/mdv507},
  publisher = {Elsevier BV},
}

@Article{Jonker2013,
  author    = {Jonker, Emile F. F. and Visser, Leonardus G. and Roukens, Anna H.},
  journal   = {Therapeutic Advances in Vaccines},
  title     = {Advances and controversies in yellow fever vaccination},
  year      = {2013},
  issn      = {2051-0136},
  month     = aug,
  number    = {4},
  pages     = {144--152},
  volume    = {1},
  doi       = {10.1177/2051013613498954},
  publisher = {SAGE Publications},
}

@Article{Prachi2012,
  author    = {Prachi, Prachi and Biagini, Massimiliano and Bagnoli, Fabio},
  journal   = {Drug Development Research},
  title     = {Vaccinology Is Turning into an Omics‐Based Science},
  year      = {2012},
  issn      = {1098-2299},
  month     = oct,
  number    = {8},
  pages     = {547--558},
  volume    = {73},
  doi       = {10.1002/ddr.21048},
  publisher = {Wiley},
}

@Article{Bhattacharya2018,
  author    = {Bhattacharya, Sanchita and Dunn, Patrick and Thomas, Cristel G. and Smith, Barry and Schaefer, Henry and Chen, Jieming and Hu, Zicheng and Zalocusky, Kelly A. and Shankar, Ravi D. and Shen-Orr, Shai S. and Thomson, Elizabeth and Wiser, Jeffrey and Butte, Atul J.},
  journal   = {Scientific Data},
  title     = {ImmPort, toward repurposing of open access immunological assay data for translational and clinical research},
  year      = {2018},
  issn      = {2052-4463},
  month     = feb,
  number    = {1},
  volume    = {5},
  doi       = {10.1038/sdata.2018.15},
  publisher = {Springer Science and Business Media LLC},
}

@Article{Edgar2002,
  author    = {Edgar, R.},
  journal   = {Nucleic Acids Research},
  title     = {Gene Expression Omnibus: NCBI gene expression and hybridization array data repository},
  year      = {2002},
  issn      = {1362-4962},
  month     = jan,
  number    = {1},
  pages     = {207--210},
  volume    = {30},
  doi       = {10.1093/nar/30.1.207},
  publisher = {Oxford University Press (OUP)},
}

@Article{Weinstein2013,
  author    = {Weinstein, John N and Collisson, Eric A and Mills, Gordon B and Shaw, Kenna R Mills and Ozenberger, Brad A and Ellrott, Kyle and Shmulevich, Ilya and Sander, Chris and Stuart, Joshua M},
  journal   = {Nature Genetics},
  title     = {The Cancer Genome Atlas Pan-Cancer analysis project},
  year      = {2013},
  issn      = {1546-1718},
  month     = sep,
  number    = {10},
  pages     = {1113--1120},
  volume    = {45},
  doi       = {10.1038/ng.2764},
  publisher = {Springer Science and Business Media LLC},
}

@Article{Liyanage2024,
  author    = {Liyanage, Jayamini C and Prendergast, Luke and Staudte, Robert and De Livera, Alysha M},
  journal   = {Bioinformatics},
  title     = {MetaHD: a multivariate meta-analysis model for metabolomics data},
  year      = {2024},
  issn      = {1367-4811},
  month     = jul,
  number    = {7},
  volume    = {40},
  doi       = {10.1093/bioinformatics/btae470},
  editor    = {Wren, Jonathan},
  publisher = {Oxford University Press (OUP)},
}

@Article{Meyners2007,
  author    = {Meyners, Michael},
  journal   = {Food Quality and Preference},
  title     = {Least equivalent allowable differences in equivalence testing},
  year      = {2007},
  issn      = {0950-3293},
  month     = apr,
  number    = {3},
  pages     = {541--547},
  volume    = {18},
  doi       = {10.1016/j.foodqual.2006.07.005},
  publisher = {Elsevier BV},
}

@Article{Hagan2022,
  author    = {Hagan, Thomas and Gerritsen, Bram and Tomalin, Lewis E. and Fourati, Slim and Mulè, Matthew P. and Chawla, Daniel G. and Rychkov, Dmitri and Henrich, Evan and Miller, Helen E. R. and Diray-Arce, Joann and Dunn, Patrick and Lee, Audrey and Deckhut-Augustine, A. and Gottardo, R. and Haddad, E. K. and Hafler, D. A. and Harris, E. and Farber, D. and Kleinstein, S. H. and Levy, O. and McElrath, J. and Montgomery, R. R. and Peters, B. and Pulendran, B. and Rahman, A. and Reed, E. F. and Rouphael, N. and Sarwal, M. M. and Sékaly, R. P. and Fernandez-Sesma, A. and Sette, A. and Stuart, K. and Togias, A. and Tsang, J. S. and Levy, Ofer and Gottardo, Raphael and Sarwal, Minne M. and Tsang, John S. and Suárez-Fariñas, Mayte and Sékaly, Rafick-Pierre and Kleinstein, Steven H. and Pulendran, Bali},
  journal   = {Nature Immunology},
  title     = {Transcriptional atlas of the human immune response to 13 vaccines reveals a common predictor of vaccine-induced antibody responses},
  year      = {2022},
  issn      = {1529-2916},
  month     = oct,
  number    = {12},
  pages     = {1788--1798},
  volume    = {23},
  doi       = {10.1038/s41590-022-01328-6},
  publisher = {Springer Science and Business Media LLC},
}

@Article{Alonso2005,
  author    = {Alonso, Ariel and Molenberghs, Geert and Geys, Helena and Buyse, Marc and Vangeneugden, Tony},
  journal   = {Statistics in Medicine},
  title     = {A unifying approach for surrogate marker validation based on Prentice’s criteria},
  year      = {2005},
  issn      = {1097-0258},
  number    = {2},
  pages     = {205--221},
  volume    = {25},
  doi       = {10.1002/sim.2315},
  publisher = {Wiley},
}

@Article{Abrahantes2004,
  author    = {Abrahantes, José Cortiñas and Molenberghs, Geert and Burzykowski, Tomasz and Shkedy, Ziv and Abad, Ariel Alonso and Renard, Didier},
  journal   = {Computational Statistics and Data Analysis},
  title     = {Choice of units of analysis and modeling strategies in multilevel hierarchical models},
  year      = {2004},
  issn      = {0167-9473},
  month     = oct,
  number    = {3},
  pages     = {537--563},
  volume    = {47},
  doi       = {10.1016/j.csda.2003.12.003},
  publisher = {Elsevier BV},
}

@Article{Meyners2012,
  author    = {Meyners, Michael},
  journal   = {Food Quality and Preference},
  title     = {Equivalence tests – A review},
  year      = {2012},
  issn      = {0950-3293},
  month     = dec,
  number    = {2},
  pages     = {231--245},
  volume    = {26},
  doi       = {10.1016/j.foodqual.2012.05.003},
  publisher = {Elsevier BV},
}

@Article{Walker2010,
  author    = {Walker, Esteban and Nowacki, Amy S.},
  journal   = {Journal of General Internal Medicine},
  title     = {Understanding Equivalence and Noninferiority Testing},
  year      = {2010},
  issn      = {1525-1497},
  month     = sep,
  number    = {2},
  pages     = {192--196},
  volume    = {26},
  doi       = {10.1007/s11606-010-1513-8},
  publisher = {Springer Science and Business Media LLC},
}

@Article{Roever2020,
  author    = {Röver, Christian},
  journal   = {Journal of Statistical Software},
  title     = {Bayesian Random-Effects Meta-Analysis Using the bayesmeta R Package},
  year      = {2020},
  issn      = {1548-7660},
  number    = {6},
  volume    = {93},
  doi       = {10.18637/jss.v093.i06},
  publisher = {Foundation for Open Access Statistic},
}

@Article{Tibaldi2003,
  author    = {Tibaldi, Fabián and Abrahantes, José Cortiñas and Molenberghs, Geert and Renard, Didier and Burzykowski, Tomasz and Buyse, Marc and Parmar, Max and Stijnen, Theo and Wolfinger, Russ},
  journal   = {Journal of Statistical Computation and Simulation},
  title     = {Simplified hierarchical linear models for the evaluation of surrogate endpoints},
  year      = {2003},
  issn      = {1563-5163},
  month     = sep,
  number    = {9},
  pages     = {643--658},
  volume    = {73},
  doi       = {10.1080/0094965031000062177},
  publisher = {Informa UK Limited},
}

@Book{2005,
  author    = {Burzykowski, Tomasz and Molenberghs, Geert and Buyse, Marc},
  editor    = {Burzykowski, Tomasz and Molenberghs, Geert and Buyse, Marc},
  publisher = {Springer New York},
  title     = {The Evaluation of Surrogate Endpoints},
  year      = {2005},
  isbn      = {9780387270807},
  doi       = {10.1007/b138566},
  issn      = {1431-8776},
  journal   = {Statistics for Biology and Health},
  ranking   = {rank5},
}

@Misc{VanDerElst2014,
  author    = {Van Der Elst, Wim and Stijven, Florian and Ong, Fenny and De Witte, Dries and Deliorman, Gokce and Meyvisch, Paul and Poveda, Alvaro and Alonso, Ariel and Ensor, Hannah and Weir, Christoper and Molenberghs, Geert},
  month     = mar,
  title     = {Surrogate: Evaluation of Surrogate Endpoints in Clinical Trials},
  year      = {2014},
  doi       = {10.32614/cran.package.surrogate},
  journal   = {CRAN: Contributed Packages},
  publisher = {The R Foundation},
}

@Article{Rinchai2021,
  author    = {Rinchai, Darawan and Roelands, Jessica and Toufiq, Mohammed and Hendrickx, Wouter and Altman, Matthew C and Bedognetti, Davide and Chaussabel, Damien},
  journal   = {Bioinformatics},
  title     = {BloodGen3Module: blood transcriptional module repertoire analysis and visualization using R},
  year      = {2021},
  issn      = {1367-4811},
  month     = feb,
  number    = {16},
  pages     = {2382--2389},
  volume    = {37},
  doi       = {10.1093/bioinformatics/btab121},
  editor    = {Martelli, Pier Luigi},
  publisher = {Oxford University Press (OUP)},
}

@Article{McKenzie2024,
  author    = {McKenzie, Joanne E. and Veroniki, Areti Angeliki},
  journal   = {Journal of Clinical Epidemiology},
  title     = {A brief note on the random-effects meta-analysis model and its relationship to other models},
  year      = {2024},
  issn      = {0895-4356},
  month     = oct,
  pages     = {111492},
  volume    = {174},
  doi       = {10.1016/j.jclinepi.2024.111492},
  publisher = {Elsevier BV},
}

@Article{Efron1987,
  author    = {Efron, Bradley},
  journal   = {Journal of the American Statistical Association},
  title     = {Better Bootstrap Confidence Intervals},
  year      = {1987},
  issn      = {1537-274X},
  month     = mar,
  number    = {397},
  pages     = {171--185},
  volume    = {82},
  doi       = {10.1080/01621459.1987.10478410},
  publisher = {Informa UK Limited},
}

@Article{Koo2016,
  author    = {Koo, Terry K. and Li, Mae Y.},
  journal   = {Journal of Chiropractic Medicine},
  title     = {A Guideline of Selecting and Reporting Intraclass Correlation Coefficients for Reliability Research},
  year      = {2016},
  issn      = {1556-3707},
  number    = {2},
  pages     = {155--163},
  volume    = {15},
  doi       = {10.1016/j.jcm.2016.02.012},
  publisher = {Elsevier BV},
}
\newpage

\section{Web Appendix D: Supplementary Figures}

\begin{figure}[ht]
\centerline{\includegraphics[height = 11cm]{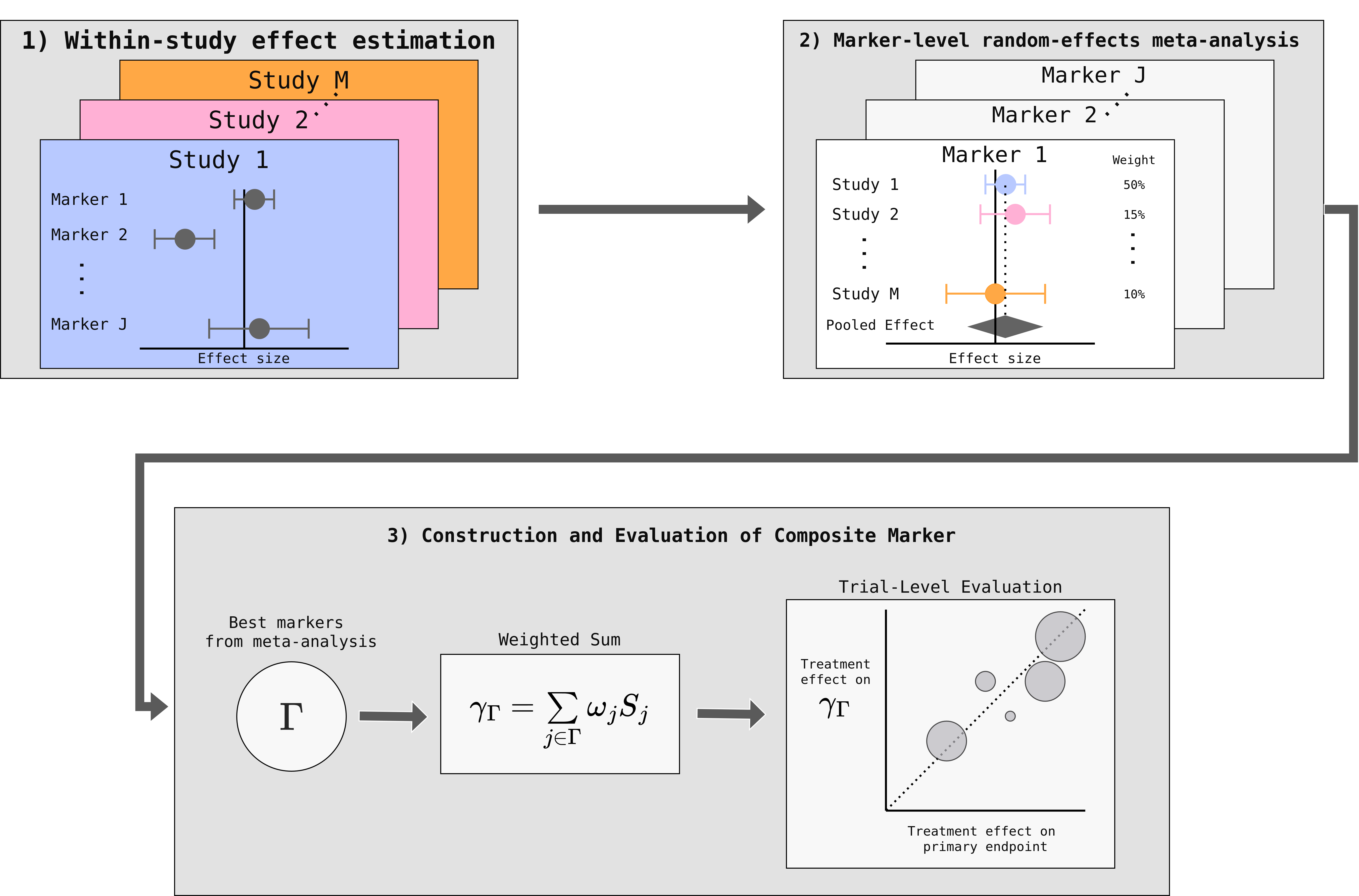}}
\caption{\textbf{Graphical overview of RISE-Meta}. \label{webfig3}}
\end{figure}

\begin{figure}[ht]
\centerline{\includegraphics[width = \textwidth]{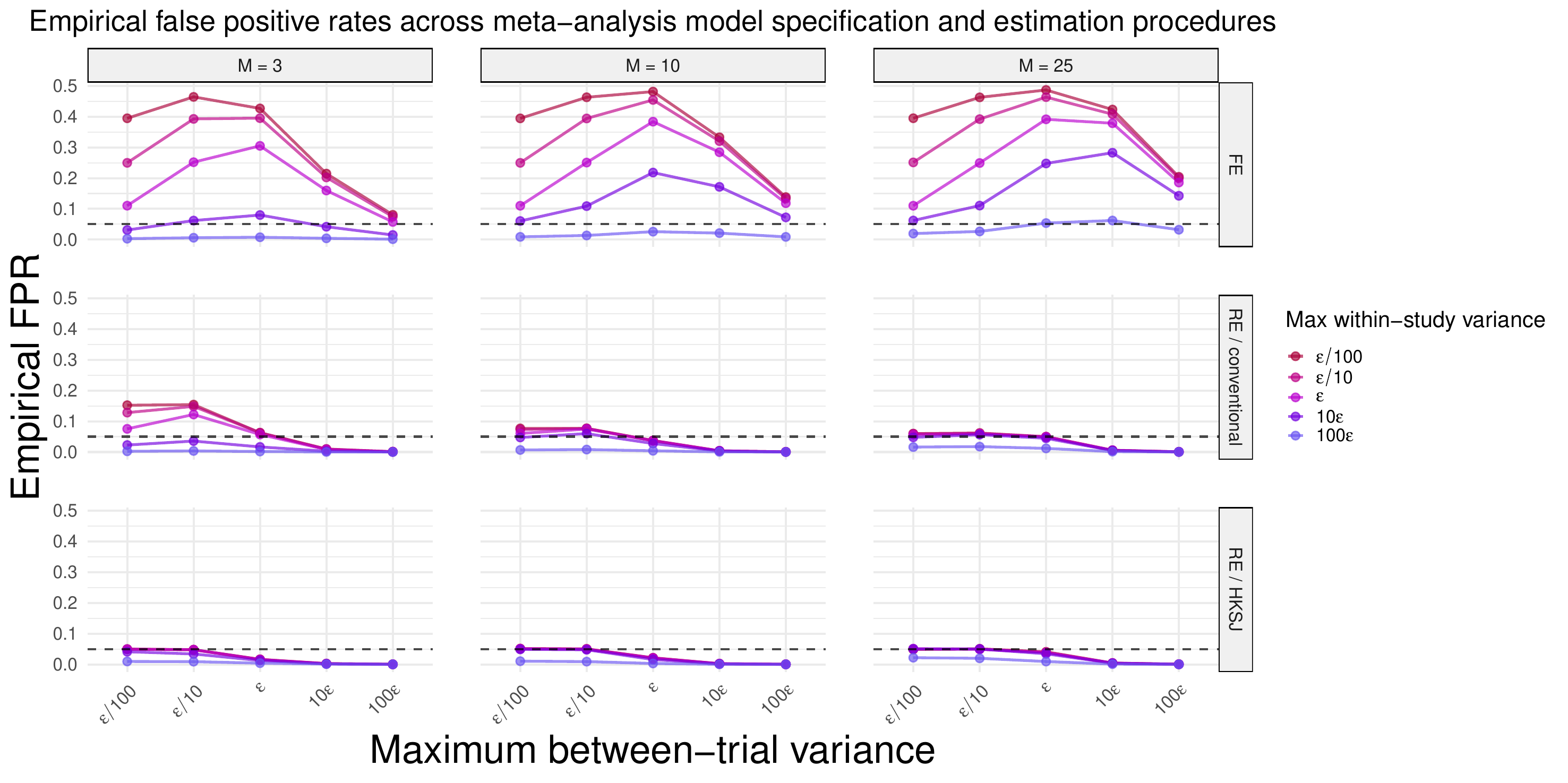}}
\caption{ \textbf{Empirical false positive rates for different meta-analysis model specifications and estimation procedures.} Each plot in the grid gives results for one combination of model specification/estimation procedure for the between-study variance $\tau^{2}$ and number of trials $M$. \textit{FE} and \textit{RE} refer to fixed and random-effects models. \textit{Conventional} refers to the conventional estimator of $\tau^{2}$ using the normal approximation whereas \textit{HKSJ} refers to the Hartung-Knapp-Sidik-Jonkman approach. For each plot, the empirical FPR is plotted against the maximum between-trial variance parameter $u_{\tau^{2}, max}$ where the coloured lines refer to different values of the maximum within-trial variance parameter $u_{\nu, max}$. Both of these parameters are given relative to a fixed value of $\varepsilon = 0.1$, which defines the equivalence margin for the TOST. The nominal false positive rate $\alpha = 0.05$ is plotted as a dashed line. \label{webfig4}}
\end{figure}

\begin{figure}[ht]
\centerline{\includegraphics[width = \textwidth]{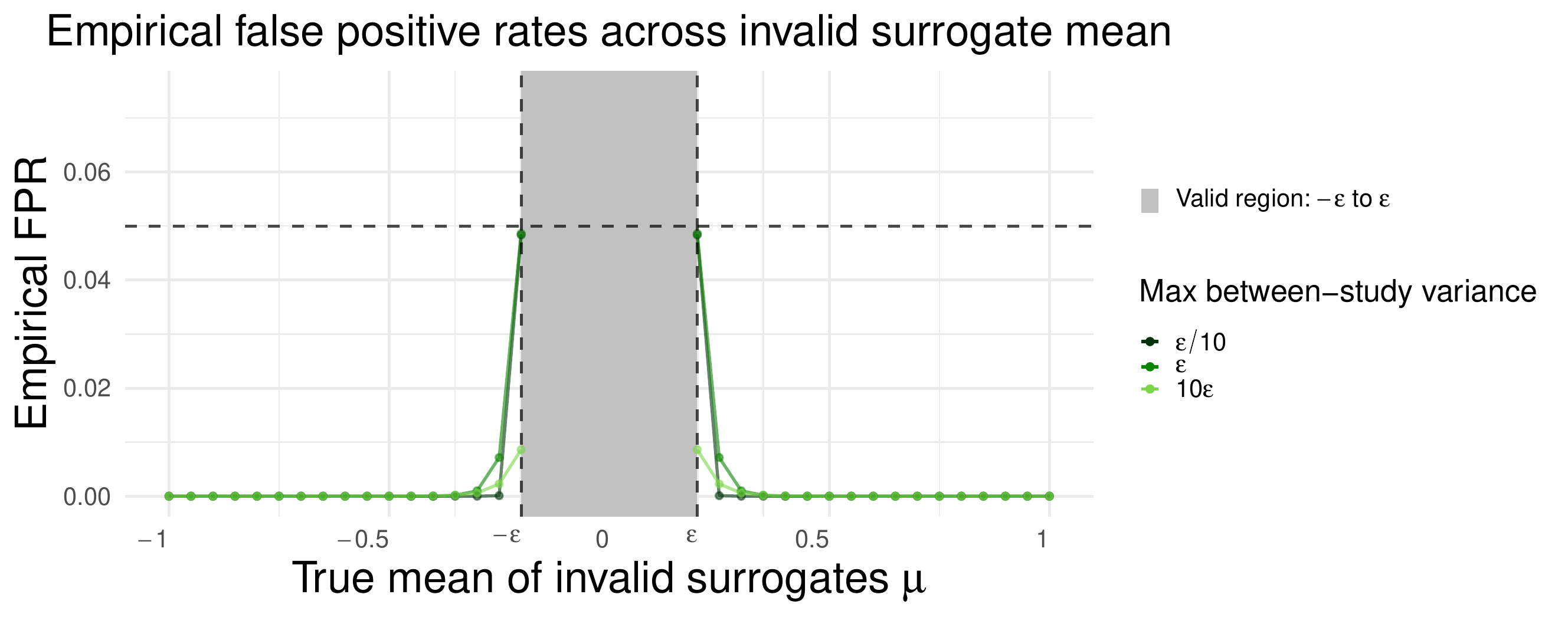}}
\caption{\textbf{Empirical false positive rates across different true mean values for invalid surrogate generation.} The grey area denotes the valid region $(-\varepsilon, \varepsilon)$, whereas the rest corresponds to the invalid region. True mean generation on these boundaries corresponds to the least favourable configuration investigated in other simulations. The line colours correspond to different values of $u_{\tau^{2}, max}$ determining the maximum between-study variance. The fixed parameter values were $u_{\tau^{2}, max} = \varepsilon/10$, $\alpha = 0.05$, $M = 25$, $n_{m} = 250$, and $\varepsilon = 0.2$. \label{webfig5}}
\end{figure}

\begin{figure}[ht]
\centerline{\includegraphics[width = \textwidth]{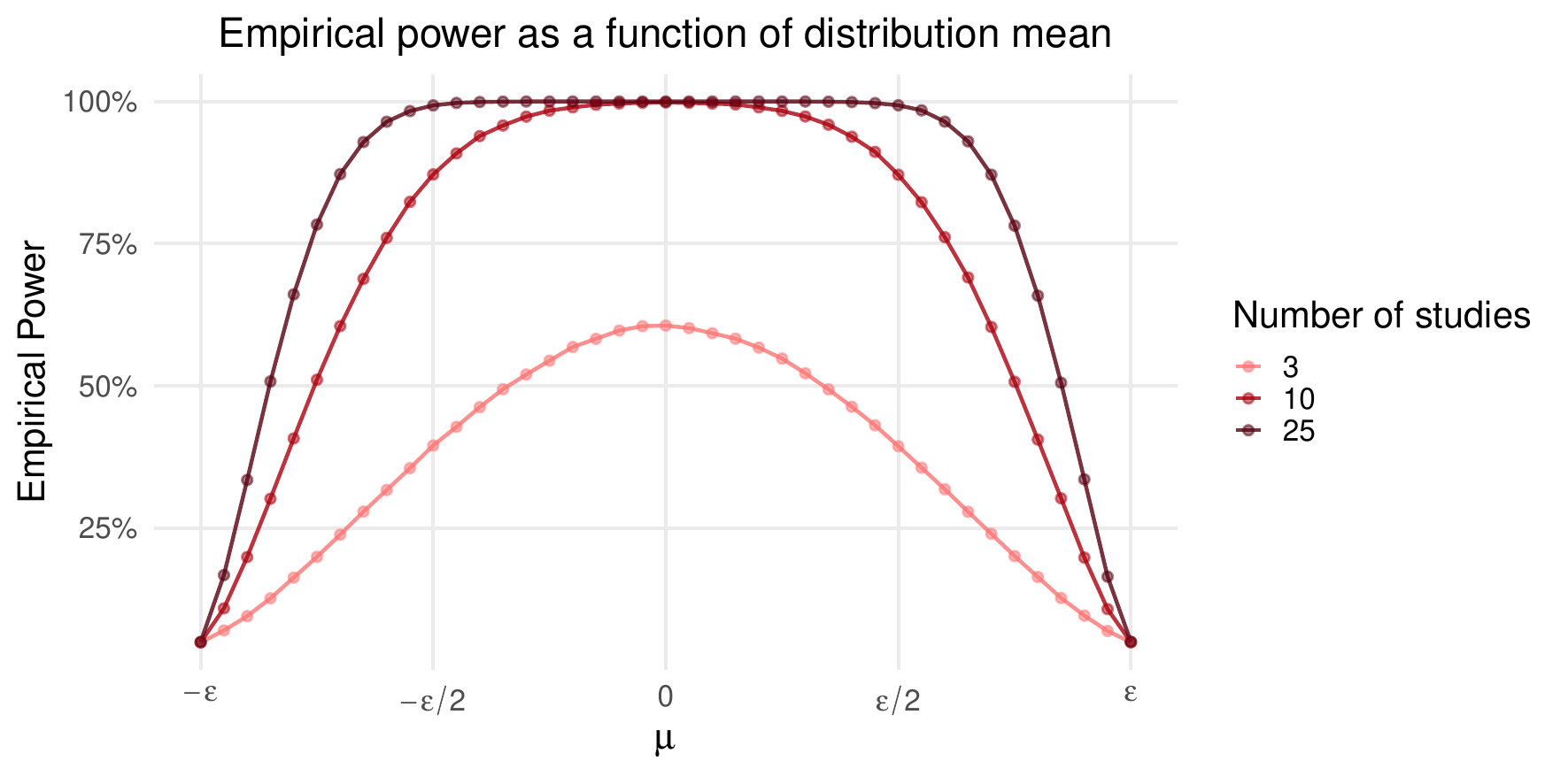}}
\caption{\textbf{Empirical power across different true mean values for valid surrogate generation.} The x axis corresponds to different values in the valid region $(-\varepsilon, \varepsilon)$ for true mean generation. The line colours correspond to different values of the number of studies $M$. The fixed parameter values were $u_{\tau^{2}, max} = u_{\nu, max} = \varepsilon/10$, $\alpha = 0.05$, $n_{m} = 50$, and $\varepsilon = 0.2$. \label{webfig6}}
\end{figure}

\begin{figure}[ht]
\centerline{\includegraphics[width = \textwidth]{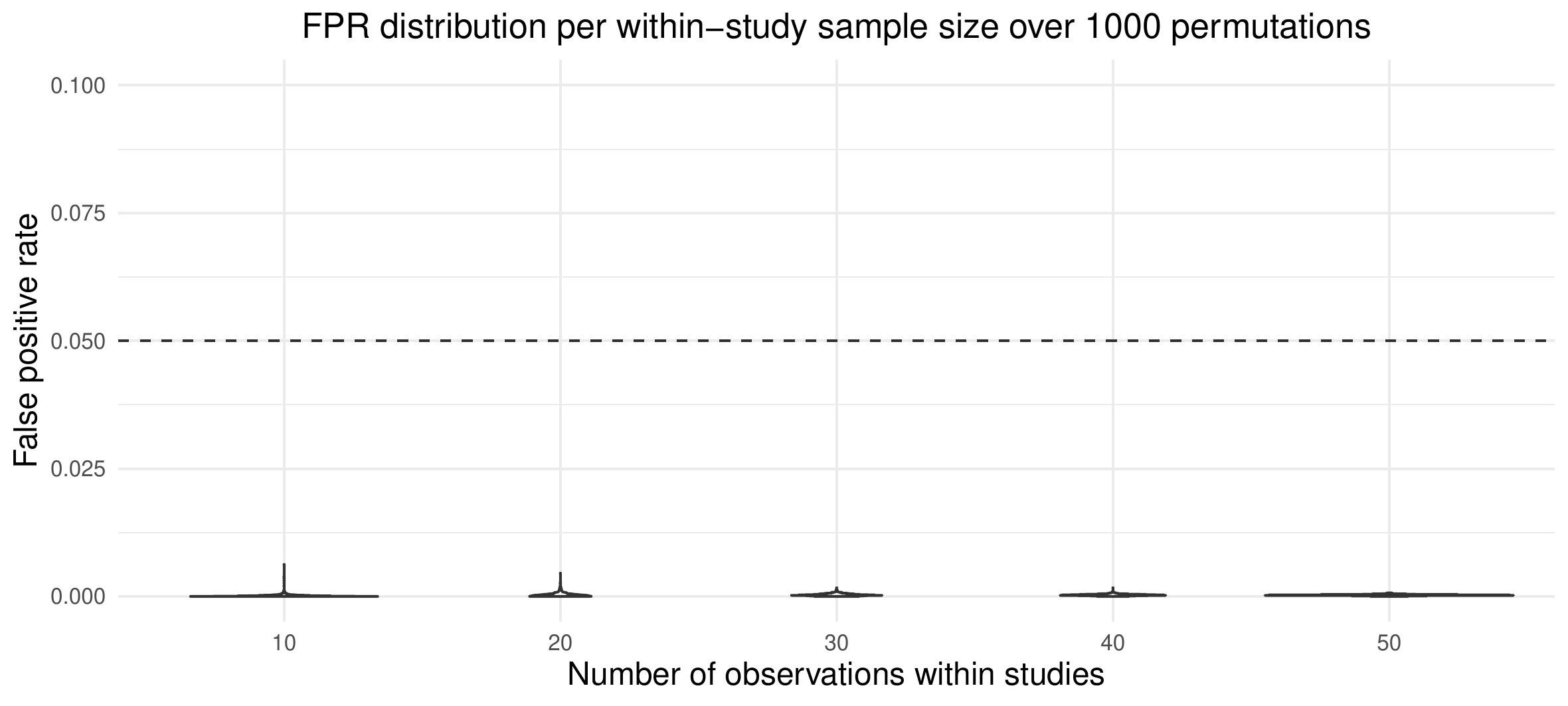}}
\caption{\textbf{Nonparametric simulation investigating the impact of the number of samples within each study.} The mean empirical false positive rates across 1,000 permuted datasets is shown as violin plots. Each value on the x-axis corresponds to the number of subsampled observations for a fixed pool of 6 studies containing at least 50 observations. The horizontal dashed line corresponds to the nominal false positive rate $\alpha = 0.05$. \label{webfig7}}
\end{figure}

\begin{figure}[ht]
\centerline{\includegraphics[width = \textwidth]{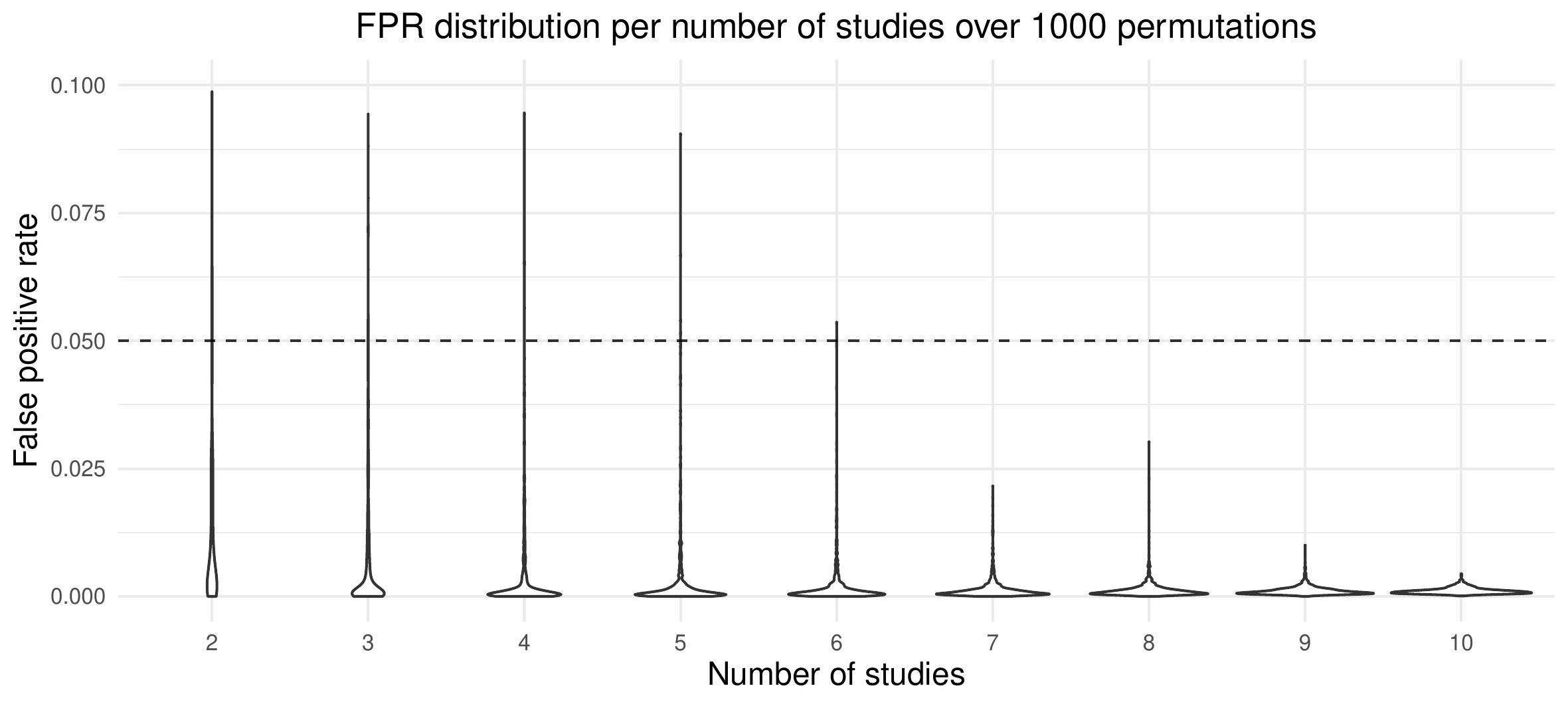}}
\caption{\textbf{Nonparametric simulation investigating the impact of the number of studies for meta-analysis.} The mean empirical false positive rates across 1,000 permuted datasets is shown as violin plots. Each value on the x-axis corresponds to the number of subsampled studies from a fixed pool of 10 studies with at least 25 individuals. The horizontal dashed line corresponds to the nominal false positive rate $\alpha = 0.05$. \label{webfig8}}
\end{figure}

\begin{figure}[ht]
\centerline{\includegraphics[width = \textwidth]{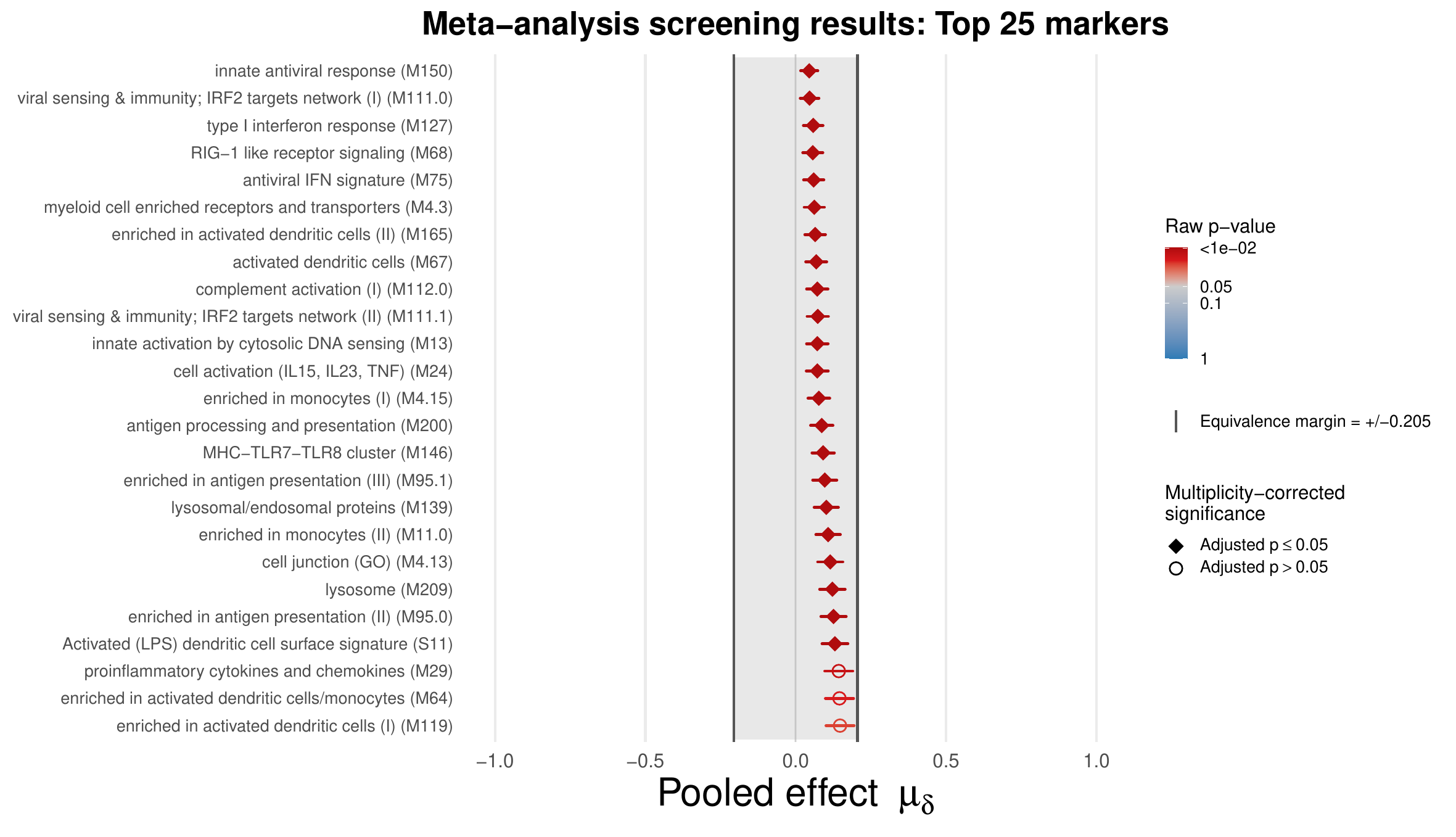}}
\caption{\textbf{Results from the marker-level meta-analysis when using a fixed-effect model.} This shows the pooled effect estimates for the top 25 markers by raw p-value. The shaded section is the equivalence region, delimited by the equivalence margins $\varepsilon \approx \pm 0.205$. The colour corresponds to the raw p-value, where shades of red indicate points whose $90\%$ CIs are contained entirely within these bounds. The shape of the point denotes the significance decision after adjusting for test multiplicity, which includes the top 16 markers. \label{webfig9}}
\end{figure}

\begin{figure}[ht]
\centerline{\includegraphics[width = \textwidth]{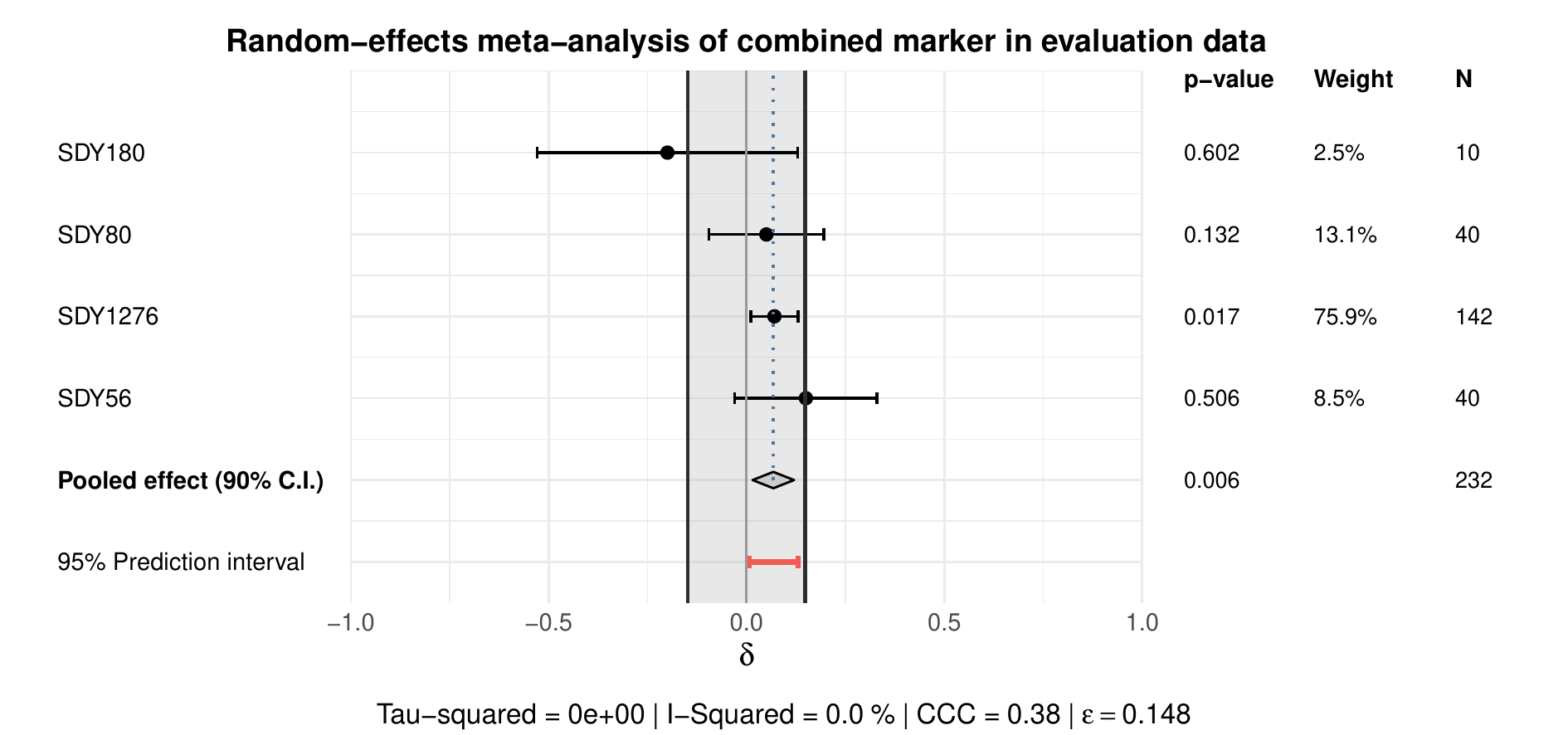}}
\caption{\textbf{Results from the evaluation of the 16-marker composite signature using a fixed-effect model.} This shows the effect estimates $\delta$ for each of the 4 studies, and the pooled effect estimate. The within-study p-values, meta-analysis weights, and number of observations $N$ are given on the right hand side. The shaded section is the equivalence region, delimited by the equivalence margins $\varepsilon \approx \pm 0.148$. The pooled effect and its $90\%$ CI are contained with this region, corresponding to a TOST p-value of $\approx 0.006$.\label{webfig10}}
\end{figure}

\begin{figure}[ht]
\centerline{\includegraphics[width = \textwidth]{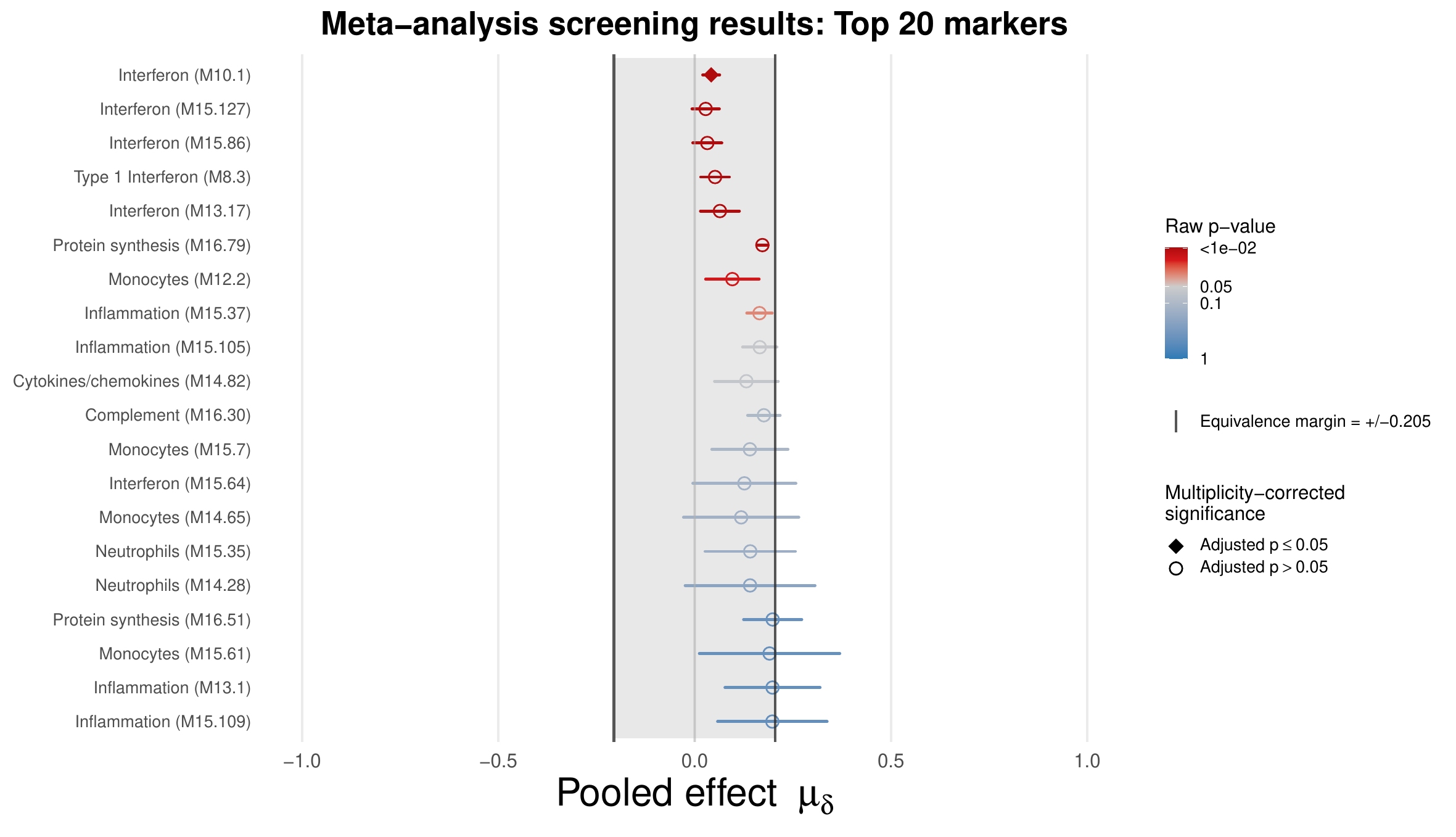}}
\caption{\textbf{Results from the marker-level meta-analysis when using the BloodGen3Modules.} This shows the pooled effect estimates for the top 20 markers by raw p-value. The shaded section is the equivalence region, delimited by the equivalence margins $\varepsilon \approx \pm 0.205$. The colour corresponds to the raw p-value, where shades of red indicate points whose $90\%$ CIs are contained entirely within these bounds. The shape of the point denotes the significance decision after adjusting for test multiplicity, which includes only the top marker. \label{webfig11}}
\end{figure}

\begin{figure}[ht]
\centerline{\includegraphics[width = \textwidth]{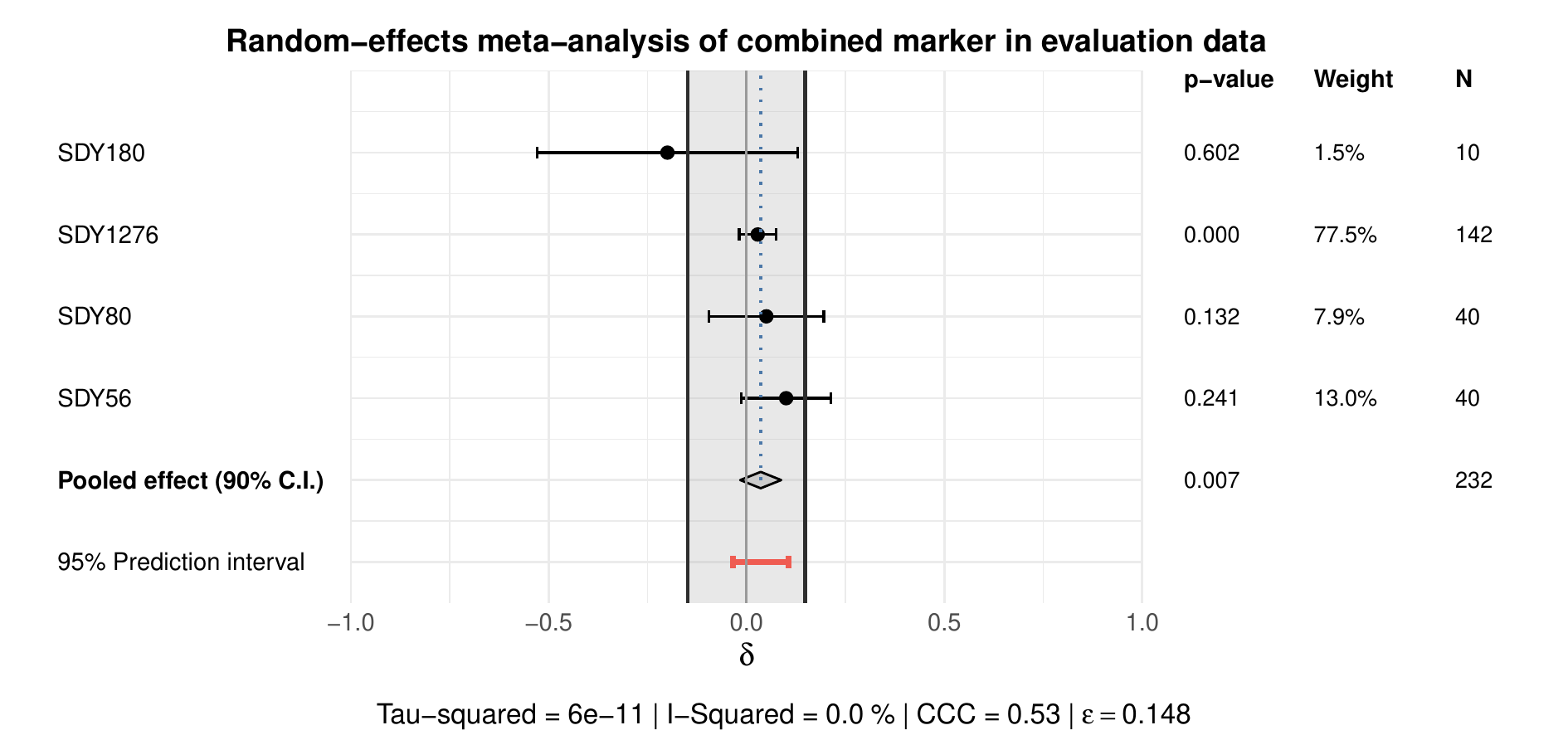}}
\caption{\textbf{Results from the evaluation of the single-marker (Interferon - M10.1) surrogate signature using the BloodGen3Modules.} This shows the effect estimates $\delta$ for each of the 4 studies, and the pooled effect estimate. The within-study p-values, meta-analysis weights, and number of observations $N$ are given on the right hand side. The shaded section is the equivalence region, delimited by the equivalence margins $\varepsilon \approx \pm 0.148$. The pooled effect and its $90\%$ CI are contained with this region, corresponding to a TOST p-value of $\approx 0.013$.\label{webfig12}}
\end{figure}

\begin{figure}[ht]
\centerline{\includegraphics[width = \textwidth]{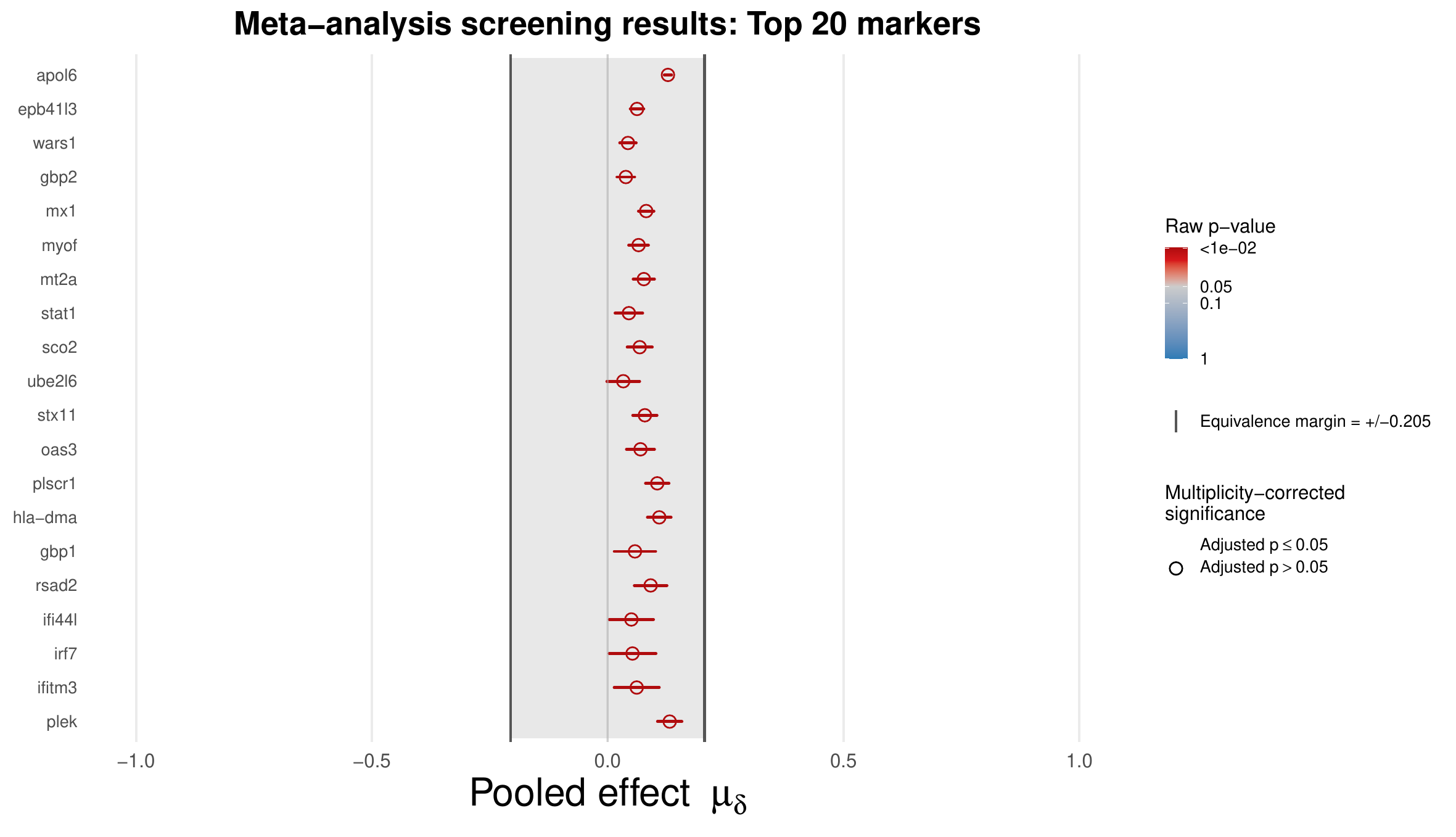}}
\caption{\textbf{Results from the marker-level meta-analysis when not aggregating to the geneset-level.} This shows the pooled effect estimates for the top 20 markers by raw p-value. The shaded section is the equivalence region, delimited by the equivalence margins $\varepsilon \approx \pm 0.205$. The colour corresponds to the raw p-value, where shades of red indicate points whose $90\%$ CIs are contained entirely within these bounds. The shape of the point denotes the significance decision after adjusting for test multiplicity, which includes no markers. \label{webfig13}}
\end{figure}

\end{refsection}

\end{document}